%% file: mfu3.tex
\documentclass[preprint,sort&compress,12pt]{elsarticle}

\input{headers/header-common.tex}

\input{headers/header-xiao.tex}


\graphicspath{ {./figs/} {./figs/jinlong/} {./figs/jianxun/}}

\journal{Elsevier}

\begin{document}

\begin{frontmatter}



  \title{A Bayesian Calibration--Prediction Method for Reducing Model-Form Uncertainties with
    Application in RANS Simulations}



\author{Jin-Long Wu\corref{corjl}}
\author{Jian-Xun Wang\corref{corjxw}}
\author{Heng Xiao\corref{corxh}}
\cortext[corxh]{Corresponding author. Tel: +1 540 231 0926}
\ead{hengxiao@vt.edu}

\address{Department of Aerospace and Ocean Engineering, Virginia Tech, Blacksburg, VA 24060, United States}

\begin{abstract}
  Model-form uncertainties in complex mechanics systems are a major obstacle for predictive
  simulations. Reducing these uncertainties is critical for stake-holders to make risk-informed
  decisions based on numerical simulations.  For example, Reynolds-Averaged Navier-Stokes (RANS)
  simulations are increasingly used in the design, analysis, and safety assessment of
  mission-critical systems involving turbulent flows. However, for many practical flows the RANS
  predictions have large model-form uncertainties originating from the uncertainty in the modeled
  Reynolds stresses.  Recently, a physics-informed Bayesian framework has been proposed to quantify
  and reduce model-form uncertainties in RANS simulations for flows by utilizing sparse observation
  data.  However, in the design stage of engineering systems, when the system or device has not been
  built yet, measurement data are usually not available.  In the present work we extend the original
  framework to scenarios where there are no available data on the flow to be predicted. In the
  proposed method, we first calibrate the model discrepancy on a related flow with available data,
  leading to a statistical model for the uncertainty distribution of the Reynolds stress
  discrepancy.  The obtained distribution is then sampled to correct the RANS-modeled Reynolds
  stresses for the flow to be predicted. The extended framework is a Bayesian
  calibration--prediction method for reducing model-form uncertainties. The merits of the proposed
  method are demonstrated on two flows that are challenging to standard RANS models.  By not
  requiring observation data on the flow to be predicted, the present calibration--prediction method
  will gain wider acceptance in practical engineering design and analysis compared to the original
  framework. While RANS modeling is chosen to demonstrate the merits of the proposed framework, the
  methodology is generally applicable to other complex mechanics models involving solids, fluids
  flows, or the coupling between the two (e.g., mechanics models for the cardiovascular systems),
  where model-form uncertainties are present in the constitutive relations.
\end{abstract}


\end{frontmatter}


\section{Introduction}
\label{sec:intro}



Model-form uncertainties in complex mechanics systems are a major obstacle for predictive
simulations. Reducing these uncertainties is critical for stake-holders to make risk-informed
decisions based on numerical simulations.  For example, the Reynolds-Averaged Navier--Stokes (RANS)
simulations have been increasingly used in the design and analysis of mission-critical systems
involving turbulent flows, thanks to the sustained growth of computational resources in the past
decades. Examples include gas turbines, hydraulic turbines, aircrafts, and more recently the
thermohydraulic system of nuclear power plants~\cite{Scheuerer:2005cu,bieder03}. Although RANS
solvers are more accurate than the state-of-the-art empirical methods and low-fidelity
models~\cite[e.g.,][]{relap7-manual}, their predictions still have significant uncertainties.  For
many practical flows in engineering, the uncertainty in the turbulence closure model embedded in the
RANS equations is the dominant source for the uncertainties in the predicted Quantities of Interest
(QoIs).  This is referred to as model-form uncertainty, which is arguably the most challenging uncertainty to
quantify in RANS models and in other complex mechanics models.

Several approaches that are generally applicable to different applications have been used to
quantify the model-form uncertainties in RANS simulations.  A widely adopted method in the
engineering communities is the parametric perturbation or model ensemble approach, in which the
baseline simulations are repeated by perturbing the coefficients in the turbulence model or by using
several turbulence models. The scattering in the obtained ensemble is used as an empirical
indication of the prediction uncertainty. However, this ad hoc approach tends to underestimate the
true uncertainties, since different models usually share similar assumptions (e.g., the Boussinesq
assumption) and thus have similar biases.  The Bayesian calibration method of Kennedy and
O'Hagan~\cite{kennedy2001bayesian} is a widely used approach for quantifying model-form
uncertainties. In their framework the discrepancy in the predicted QoI is modeled as a Gaussian
process, whose hyperparameters are calibrated with data.  However, this physics-neutral framework
treats the numerical model as a black-box, which leads to inefficient use of calibration data and
makes it difficult for the users to specify prior knowledge based on their insights of the model and
the problem.  In the past few years, some researchers have proposed open-box approaches to quantify
the model-form uncertainties in RANS simulations, which are in contrast to the general-purpose,
physics-neutral approaches described above.  Dow and Wang~\cite{dow11quanti} modeled the true eddy
viscosity field as a Gaussian process, and used Direct Numerical Simulation (DNS) data to infer its
hyperparameters and to find the optimal mean field that minimizes the misfit.  Iaccarino and
co-workers~\cite{emory2011modeling,emory2013modeling,gorle2013framework} introduced smooth
perturbations on the Reynolds stress towards several limiting states in the physically realizable
range~\cite{banerjee2007presentation,tennekes1972first}. Oliver et al.~\cite{oliver2009uncertainty}
introduced discrepancy to the Reynolds stress tensor, and modeled the discrepancy term with a
stochastic differential equation.  The essential differences among the three open-box approaches lie
in the treatment of physical prior knowledge and calibration data.  Dow and Wang~\cite{dow11quanti}
used DNS data to reduce uncertainties, resulting in a calibration--prediction procedure. In
contrast, Iaccarino et al.~\cite{emory2011modeling,emory2013modeling,gorle2013framework} and Oliver
et al.~\cite{oliver2009uncertainty} focused on uncertainty quantification and propagation by
exploiting the physical insight on the Reynolds stresses, and they did not directly use data to
reduce uncertainties.

More recently, Xiao et al.~\cite{xiao-mfu} proposed an open-box, physics-informed, Bayesian
framework for quantifying and reducing model-form uncertainties in RANS simulations. Uncertainties
are introduced to the Reynolds stresses and are parameterized compactly with physical realizability
and spatial smoothness guaranteed~\cite{oliver2009uncertainty, emory2011modeling}.  A Bayesian
inference procedure based on an iterative ensemble Kalman method~\cite{iglesias2013ensemble} is used
to quantify and reduce the uncertainties by incorporating observation data. This method combines all
sources of available information in a Bayesian framework including physical constraints, empirical
knowledge, and observation data. In scenarios where observation data (e.g., sparse velocity
measurements) are available, this framework provides a powerful method of predicting the whole-field
velocity and other QoIs with quantified uncertainties.

However, in the design stage of engineering systems or devices, the target configuration (e.g., the
new design of an aircraft or a gas turbine) has not been built yet, and thus measurement data are
not available~\cite{torenbeekdesign,mcmasters2004}. The framework of Xiao et al.~\cite{xiao-mfu} 
is not directly applicable to these
scenarios.  Without observation data the framework would degenerate to an uncertainty estimation
procedure, and the potential of the Bayesian framework cannot be fully exploited. Therefore, in the
present work we extend the original framework to predict flows with no observation data by
utilizing data from a different yet related flow.  Specifically, we first build a statistical model
for the Reynolds stress discrepancy on a related flow by incorporating available data (i.e., infer
its uncertainty distribution), and then we sample the inferred uncertainty distribution to predict
the flow of interest. The extended framework is essentially a Bayesian calibration--prediction
procedure.  We consider two typical scenarios where observation data could be available on a related
flow.  (1) When mission-critical systems are developed, laboratory experiments are often performed
on a down-scaled, geometrically similar model. Unfortunately, the experiments often have to be
performed at a reduced Reynolds number than in the prototype, possibly due to limitations of the
facilities or to avoid violating other similitudes (e.g., Mach number similarity).  Although the
Reynolds number similitude is not achieved, the experimental data obtained in the laboratory scale
model can be valuable for quantifying uncertainties in the RANS-predicted Reynolds stresses.  (2)
Industrial product developments often consist of small incremental modifications of existing, well
tested products, e.g., when a new gas turbine model is added to a family of existing products. In
this case it is possible that measurement data, obtained from either experiments or field
operations, are available on flows in a slightly different geometry from the flow to be predicted.


In the calibration--prediction framework outlined above, a pivotal assumption is that the Reynolds
stress discrepancies in the two flows, the one used for calibration and the one to be predicted, can
be described by the same statistical model. This assumption allows the extrapolation of Reynolds
stress discrepancies from the calibration case to the prediction case, where ``extrapolation''
should be interpreted in a \emph{statistical sense}.  The extrapolation of the Reynolds stress
discrepancy is justified if the two cases share overall physical characteristics despite the
differences in their specific flow conditions (e.g., Reynolds number and geometry).  The same
assumption is the basis of many wind tunnel experiments in which it is infeasible to achieve the
prototype Reynolds number with the laboratory setup.

The objective of the present work is to explore the feasibility of such a calibration--prediction
procedure.  The performance of the proposed extension is evaluated on two canonical flows, the flow
over periodic hills and the flow in a square duct, at several Reynolds
numbers. In both cases the Reynolds stress discrepancies are calibrated in a lower Reynolds number
flow with sparse observation of velocities and then are used to predict flows at higher Reynolds
numbers. Moreover, we also explore the feasibility of extrapolating the Reynolds stress discrepancy
calibrated in a square duct flow to the flow in a rectangular duct.  The remaining of the paper is
organized as follows. Section~2 outlines the model-form uncertainty quantification framework
proposed in Xiao et al.~\cite{xiao-mfu} and presents the extension proposed in the current
work. Section~3 presents and discusses numerical simulation results for two canonical flows that are
challenging for standard RANS models.  Finally, Section~4 concludes the paper.

\section{Methodology of the Bayesian Calibration--Prediction Framework}
\label{sec:method}

Consider two flows that are closely related, e.g., flows in the same geometry but at a different
Reynolds number, or flows in slightly different geometries.  One flow (to be used for calibration)
has some velocity observation data available, while the other flow (to be predicted) has no
observation data.  The overall idea of the proposed calibration--prediction framework can be
summarized as follows.  Baseline RANS simulations are first performed for both the calibration flow
and the flow to be predicted, both having discrepancies in the RANS-predicted Reynolds stresses
that are \emph{a priori} unknown. In the present framework, we assume that the discrepancies in both
cases have the same statistical distributions and model them as a random field.  We first utilized
prior knowledge and observation data to build a statistical model of the discrepancy.  Subsequently,
the obtained uncertainty distribution of the discrepancy is sampled and used to correct the baseline
Reynolds stress in the prediction case, which is then propagated to uncertainties in velocities and
in other QoIs by solving the RANS equations with the corrected Reynolds stresses.

Admittedly, since flows at different Reynolds numbers can have significantly different Reynolds
stress magnitudes, direct extrapolation of the Reynolds stress discrepancy tensor field from one
flow to another is likely to be problematic. Therefore, when extrapolating the Reynolds stress
discrepancies to the flow to be predicted, we extrapolate the projections (magnitude, shape,
orientation) of the discrepancies and not the componentwise discrepancies.  Specifically, the
log-discrepancy of the magnitude and discrepancies of the shapes parameters of the Reynolds stress
tensors are extrapolated. These projections can be considered normalized quantities independent of
the magnitude of the Reynolds stresses, and thus they can be extrapolated from one flow to another.
Note that in the calibration of Reynolds stress discrepancies, the uncertainty space is also parameterized on
the physical projections to guarantee the physical realizability (i.e., Reynolds stress tensor must
correspond to a physically possible state~\cite{tennekes1972first}).

\begin{figure}[!htbp]
  \centering
  \includegraphics[width=0.8\textwidth]{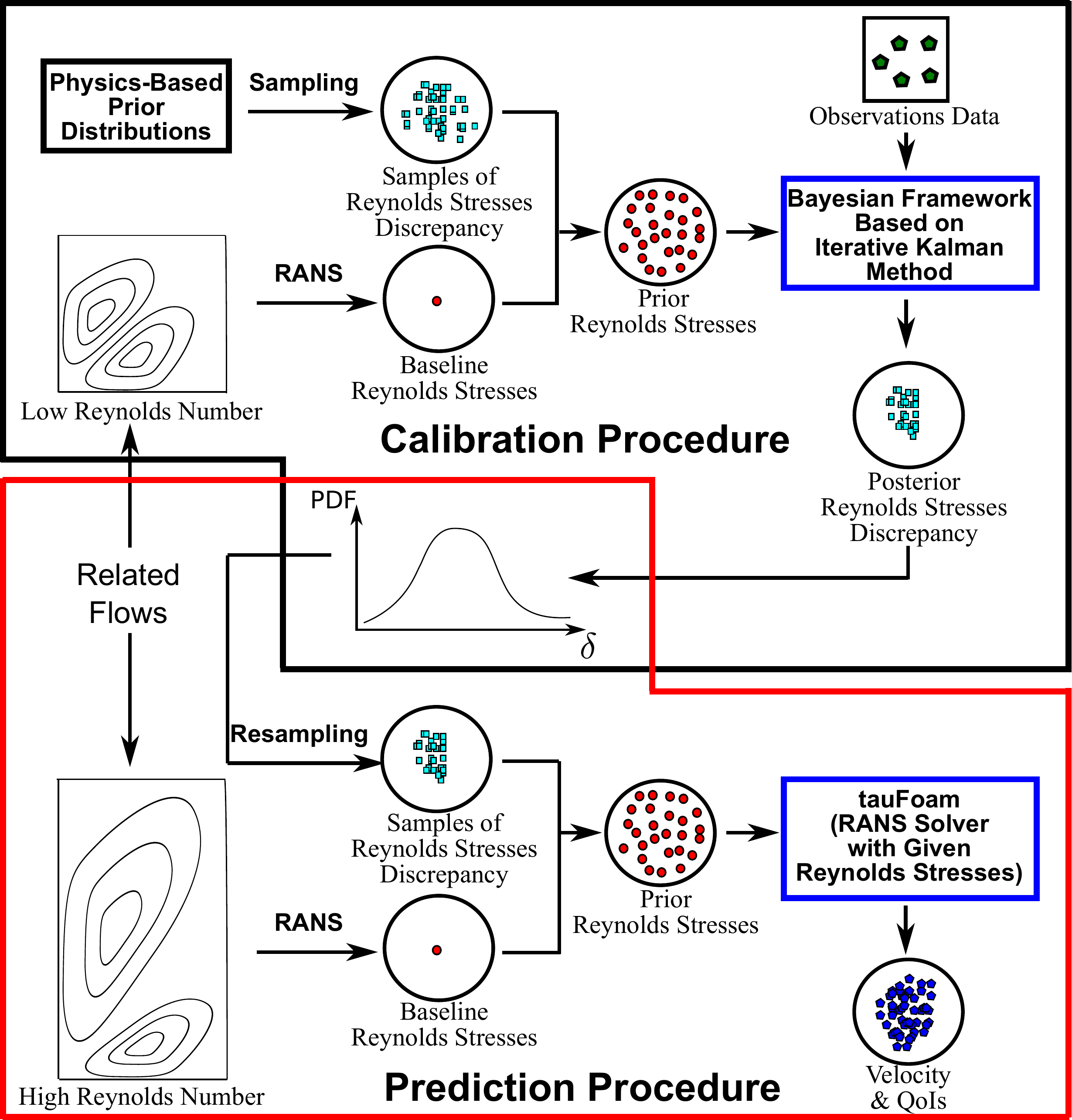}
  \caption{Schematic overview of the calibration--prediction procedure for the quantification,
    reduction, and propagation of uncertainties in RANS models. An example is used to illustrate the
    overall algorithm. In this example the uncertainty distribution of the Reynolds stress
    discrepancy is calibrated in a square duct flow at a low Reynolds number, and then the
    calibrated distribution is used to predict the flow in a rectangular duct at a higher Reynolds
    number where no data are available.}
  \label{fig:overview}
\end{figure}

In Fig.~\ref{fig:overview} we summarize the probabilistic calibration--prediction framework for RANS
modeling with uncertainty quantifications . The two components of the framework, calibration and
prediction, are introduced in Sections~\ref{sec:method-cali} and~ \ref{sec:extrap}, respectively.

\subsection{Calibration by Building a Statistical Model of the Reynolds Stress Discrepancy}
\label{sec:method-cali}

\begin{figure}[!htbp]
  \centering
   \includegraphics[width=0.9\textwidth]{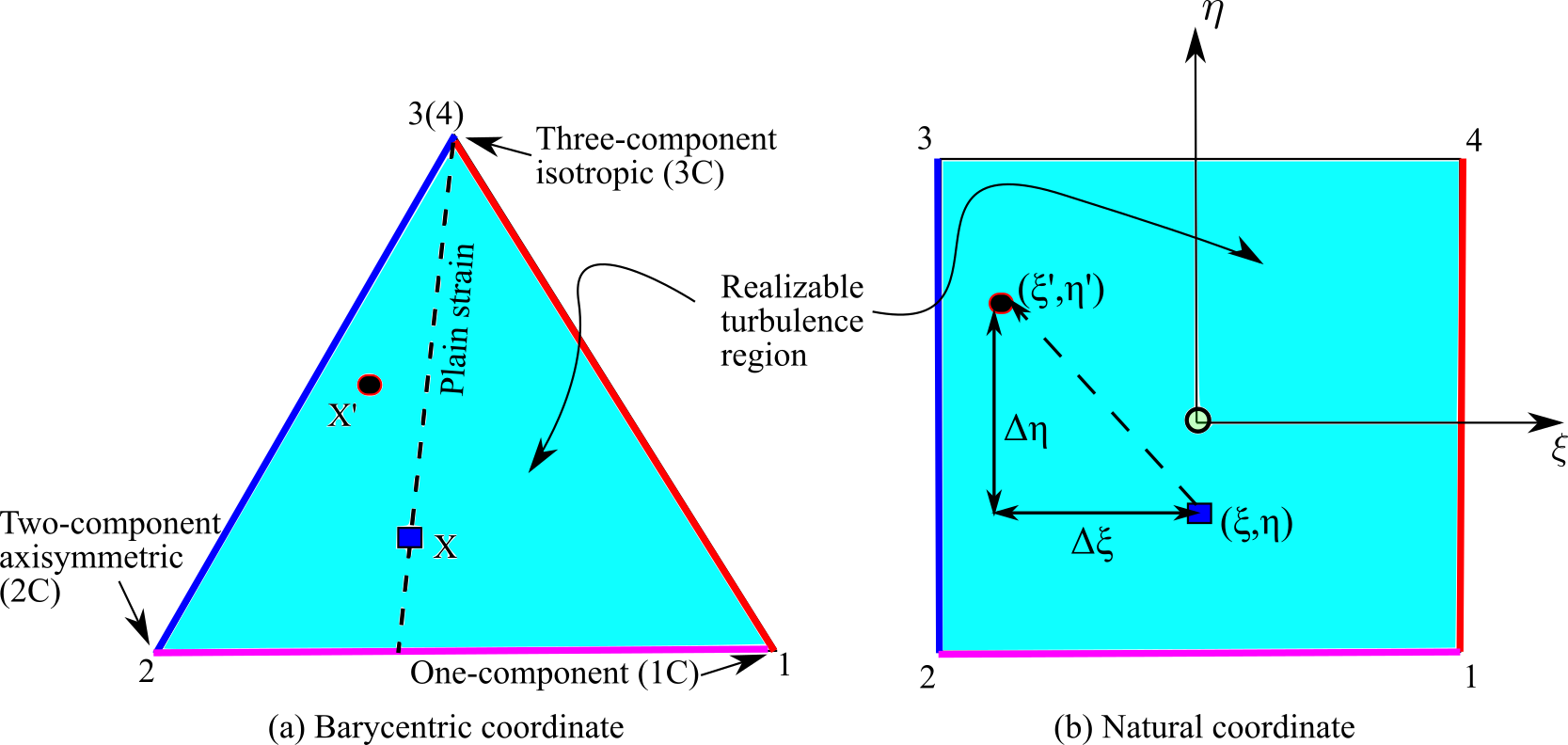}
   \caption{Mapping between the Barycentric coordinate to the natural coordinate, transforming the
     Barycentric triangle enclosing all physically realizable
     states~\cite{banerjee2007presentation,emory2013modeling} to a square through standard finite
     element shape functions. Details of the mapping can be found in the appendix of
     ref.~\cite{xiao-mfu}. Corresponding edges in the two coordinates are indicated with matching
     colors.}
  \label{fig:bary}
\end{figure}

The calibration procedure uses the method of Xiao et al.~\cite{xiao-mfu} for quantifying and
reducing model-form uncertainties. The procedure is briefly summarized here for completeness.  In
the framework, the true Reynolds stress $\bstau(x)$ is modeled as a random tensorial field with the
spatial coordinate $x$ as index and the RANS-predicted Reynolds stress $\bstaurans(x)$ as prior
mean.  To ensure physical realizability, uncertainties are injected to the physically meaningful
projections of the Reynolds stress tensor, which is transformed as follows~\cite{emory2011modeling}:
\begin{equation}
  \label{eq:tau-decomp}
  \boldsymbol{\tau} = 2 k \left( \frac{1}{3} \mathbf{I} +  \mathbf{a} \right)
  = 2 k \left( \frac{1}{3} \mathbf{I} + \mathbf{V} \Lambda \mathbf{V}^T \right)
\end{equation}
where $k$ is the turbulent kinetic energy which indicate the magnitude of $\bstau$; $\mathbf{I}$ is
the second order identity tensor; $\mathbf{a}$ is the anisotropy tensor; 
$\mathbf{V} = [\mathbf{v}_1, \mathbf{v}_2, \mathbf{v}_3]$, and $\Lambda = \textrm{diag}[\lambda_1, \lambda_2, \lambda_3]$ with
$\lambda_1+\lambda_2+\lambda_3=0$ are the orthonormal eigenvectors and eigenvalues of $\mathbf{a}$,
respectively, indicating the shape and orientation of $\bstau$. After the decomposition, the
eigenvalues $\lambda_1$, $\lambda_2$, and $\lambda_3$ are mapped to a Barycentric coordinates $(C_1,
C_2, C_3)$ with $C_1 + C_2 + C_3 = 1$. Consequently, all physically realizable states are enclosed
in the Barycentric triangle shown in Fig.~\ref{fig:bary}a. To facilitate parameterization, the
Barycentric coordinates are further transformed to the natural coordinates $(\xi, \eta)$ with the
triangle mapped to the square as shown in Fig.~\ref{fig:bary}b.  After the mapping, uncertainties are
introduced to the mapped quantities $k$, $\xi$, and $\eta$ by adding discrepancy terms to the
corresponding RANS predictions, i.e.,
\begin{subequations}
    \label{eq:delta-def}
  \begin{alignat}{2}
    \log k(x) & = &\ \log \tilde{k}^{rans}(x)  & + \delta^k(x)  \label{eq:kdelta} \\
    \xi (x) & = &\ \tilde{\xi}^{rans}(x) & + \delta^\xi(x)  \\
    \eta(x) & = &\ \tilde{\eta}^{rans}(x) & + \delta^\eta(x)
  \end{alignat}
\end{subequations}
where $\delta^k(x)$ is the log-discrepancy of the turbulent kinetic energy; $\delta^\xi(x)$ and
$\delta^\eta(x)$ are discrepancies of the Reynolds stress shape parameters $\xi$ and $\eta$,
respectively. To avoid instability due to potential reverse diffusions, uncertainties are not
introduced to the orientation ($\mathbf{v}_1, \mathbf{v}_2, \mathbf{v}_3$) of the Reynolds stresses.

Another piece of prior information is the smooth spatial distribution of the Reynolds stresses. The
smoothness is guaranteed by representing the Reynolds stress discrepancy fields $\delta^k$,
$\delta^\xi$, and $\delta^\eta$ (generically denoted as $\delta$ below) with smooth basis
functions.  Specifically, the prior distributions of the discrepancies are chosen as nonstationary
zero-mean Gaussian random fields $\mathcal{GP}(0, K)$ (also know as Gaussian processes), and the
basis set is chosen as the eigenfunctions of the kernel $K$~\cite{le2010spectral}. This choice of
basis function leads to the Karhunen--Loeve (KL) expansions of the random field. That is, the
discrepancies can be represented as follows:
\begin{equation}
  \label{eq:delta-proj}
  \delta(x, \theta) = \sum_{i=1}^\infty \omega_{i} |_{\theta_i} \; \phi_i (x) , 
\end{equation}
where $\theta$ is the realized outcome, and the coefficients $\omega_{i}$ (denoting $\omega^k_{i}$,
$\omega^{\xi}_{i}$, and $\omega^\eta_{i}$ for discrepancy fields $\delta^k$, $\delta^\xi$, and
$\delta^\eta$, respectively) are independent standard Gaussian random variables. In practice the
infinite series is truncated to $m$ terms with $m$ depending on the smoothness of the kernel $K$.

With the projections above, the discrepancies are parameterized by the coefficients $\omega^k_{i},
\, \omega^\xi_{i}, \, \omega^\eta_{i}$ with $i = 1, 2, \cdots, m$. The uncertainty distributions of
the coefficients are then inferred by using an iterative ensemble Kalman
method~\cite{iglesias2013ensemble,evensen2009data}. In this method the prior distribution of
Reynolds stresses (as parameterized by the coefficients) is first represented with samples drawn
from the distribution.  The collection of the samples, referred to as prior ensemble, is propagated
to velocities by using a forward RANS solver.  The Kalman filtering procedure is used to incorporate
velocity observation data to the prediction, yielding a corrected ensemble.  The procedure is
repeated until statistical convergence is achieved. The converged posterior ensemble is a sample
based representation of the uncertainty distributions of the Reynolds stresses and other QoIs.

The forward RANS solver \texttt{tauFoam} used in the Bayesian inference above basically computes
velocities from a \emph{given Reynolds stress field}. It is developed based on a conventional
steady-state RANS solver in OpenFOAM~\cite{openfoam} by replacing the turbulence modeling component
(i.e., solution of the transport equations for turbulent quantities) with a supplied Reynolds stress
field.

\subsection{Prediction for Flows without Observation Data with Corrected  Reynolds Stresses}
\label{sec:extrap}

The calibration procedure outlined in Section~\ref{sec:method-cali} builds a statistical model by
inferring the uncertainty distribution of the Reynolds stress discrepancy, which is represented by
the posterior ensemble obtained from the iterative Kalman method.  Next, we use the obtained
statistical model of the Reynolds stresses discrepancy to make predictions on flows with no
observation data.  To this end, we extrapolate the Reynolds stress discrepancy from the calibration
case to the prediction case, and then we use the forward RANS solver to propagate the samples of
corrected Reynolds stresses to the corresponding velocities and other QoIs.

\subsubsection{Extrapolation  to flow at a different Reynolds number}

After calibrating the uncertainty distribution of the Reynolds stress discrepancy, it is
straightforward to extrapolate to a flow with geometric similarity but not dynamic similarity (i.e.,
at a different Reynolds number).  Here we merely emphasize two potential pitfalls as mentioned
above. First, the analyst must ensure that the two flows indeed have the same overall
characteristics. If the calibration case is a flow over an airfoil without separation while the flow
to be predicted has a massive separation, the extrapolation can lead to incorrect results even
though the two cases have the same geometry. Second, it is the discrepancies of the physical
projections of the Reynolds stress that are extrapolated. Since these are also the discrepancies
inferred in the calibration procedure, it does not pose any practical difficulties.

\subsubsection{Extrapolation to flow in a different geometry}
\label{sec:extra-geo}

When the calibration case and the prediction case differ not only in Reynolds number but also in
geometry, extrapolation of the Reynolds stress discrepancies poses additional challenges. That is,
how to map a field from one geometry (e.g., a square) to another (e.g., a rectangle). The mapping
scheme depends on the physical characteristics of the two flows. The choice is inevitably
case-specific and relies on the judgment of the analyst.

\begin{figure}[!htbp]
  \centering
  \includegraphics[width=0.9\textwidth]{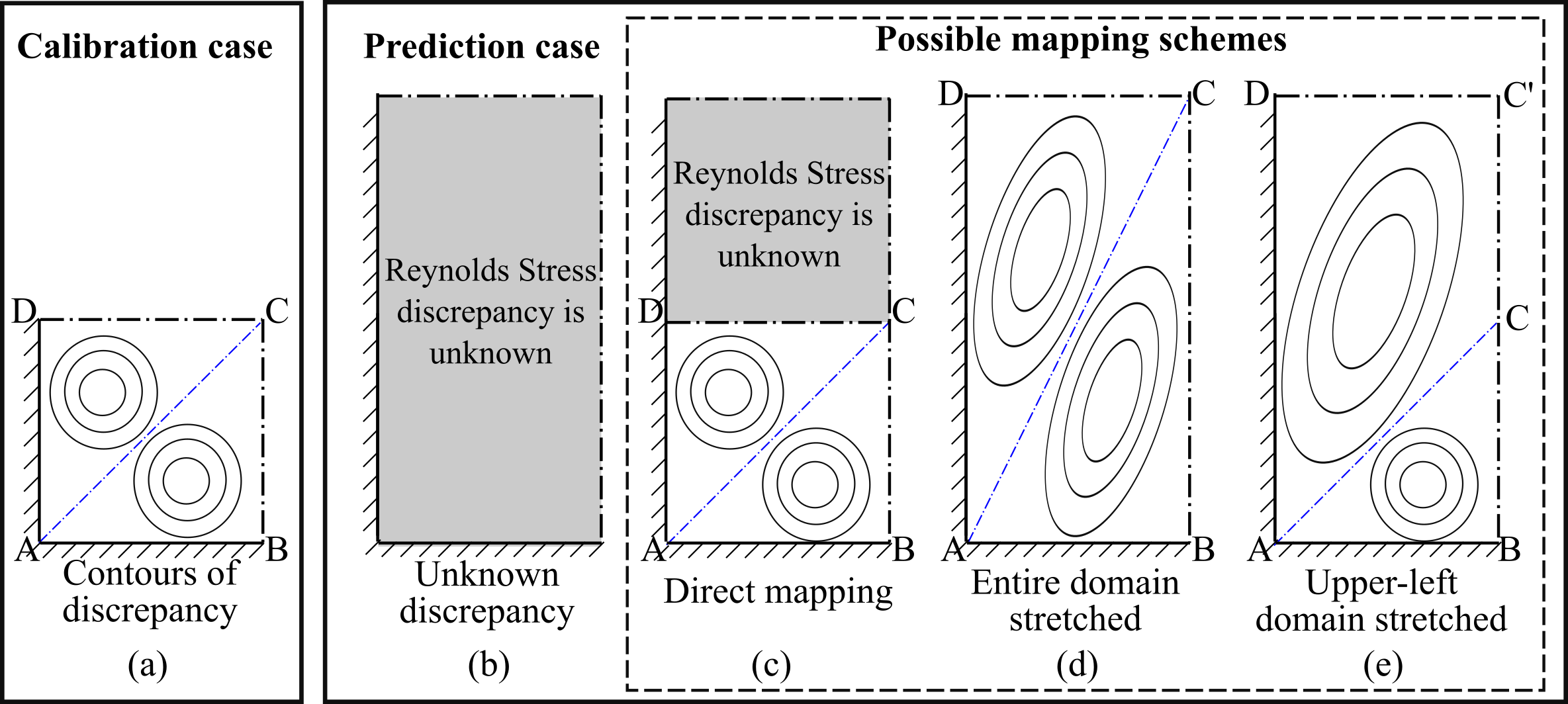}
  \caption{Schematic illustration of possible schemes for mapping the calibrated Reynolds stress
    discrepancies on a square duct to a rectangular duct, which is the flow to be
    predicted. Dash-dotted lines indicate lines of symmetry. Contours indicate Reynolds stress
    discrepancies and underlying flow structures.}
  \label{fig:ext-geometry}
\end{figure}

In Fig.~\ref{fig:ext-geometry} we use an example to illustrate the mapping scheme, where the
Reynolds stress discrepancy calibrated on the flow in a square duct is used to predict the flow in a
rectangular duct. Results for this case will be presented in Section~\ref{sec:res-geometry}.  In the
calibration (Fig.~\ref{fig:ext-geometry}a), the two contours symmetric along the diagonal indicate
spatial distribution of Reynolds stress discrepancy due to particular features of the flow, e.g.,
vortices. The shaded region in Fig.~\ref{fig:ext-geometry}b indicates that the Reynolds stress
discrepancy in the prediction case, the flow in the rectangular geometry, is unknown before the
knowledge from the square duct case is incorporated.  There are at least three possbile schemes of
extrapolating discrepancies calibrated in the square geometry to the rectangle:
\begin{enumerate}
\item \emph{Direct mapping}. As shown in Fig.~\ref{fig:ext-geometry}c, the obtained discrepancy
  field in the square is directly mapped to the lower half of the rectangle, while the Reynolds
  stress discrepancies in the upper half are still unknown. Prior knowledge can still be specified
  on the discrepancy in the unmapped region, but the uncertainty distribution in this region does
  not reflect information from the calibration case.

  The mapping between the two geometries can be better understood by comparing the vertex numbering
  (A, B, C, D) in the square domain in Fig.~\ref{fig:ext-geometry}a and the rectangle domains in
  Fig.~\ref{fig:ext-geometry}c.
\item \emph{Stretching the entire domain}. As shown in Fig.~\ref{fig:ext-geometry}d, the calibrated
  field is mapped linearly to the entire domain of the rectangle, i.e., both contours are stretched
  in the mapped domain.
\item \emph{Direct mapping of lower right domain and stretching of upper left domain}.  As shown in
  Fig.~\ref{fig:ext-geometry}e, the lower left half of the calibrated discrepancy field (A-B-C)
  below the diagonal is directly mapped to the same triangle in the rectangular domain, and the
  upper triangle (A-C-D) is mapped to the quadrilateral (A-C-C'-D) in the rectangular domain.
\end{enumerate}

The analyst must rely on the physical understanding of the flow to be predicted (i.e., the flow in
the rectangular geometry in Fig.~\ref{fig:ext-geometry}b) to decide which mapping schemes described
above are reasonable.  Even though no data is directly available from the flow to be predicted, we
still utilizes the two ingredients of a Bayesian framework, prior knowledge and data, to make
predictions. The prior knowledge is formulated based on the relations between the flows in the
calibration and prediction cases, while the data come from the calibration case. Therefore, as with
the original framework of Xiao et al.~\cite{xiao-mfu}, the extended calibration--prediction procedure
is a Bayesian framework.

\subsubsection{Resampling the posterior ensemble}

Once we identify a scheme for mapping the discrepancies, we can sample the posterior distribution of
the discrepancy obtained in the calibration and use them to make predictions.  Sampling from a
statistical model of a random field can be achieved with established
methods~\cite[e.g.,][]{Glasserman:2004ua}.  However, the actual implementation is not
straightforward and it thus worthes a detailed discussion.  There are two obstacles in the resampling
of the posterior ensemble for Reynolds stress discrepancies. First, different components in the
random field are correlated.  Second, the distribution to be sampled is not given analytically but
represented by samples. The first problem can be solved by projecting the samples to the space
spanned by the eigenvectors of the covariance matrix of the random field. The covariance matrix can
be estimated from the ensemble. After the change of coordinates, each component in a transformed
sample field can be considered realization from an independent random variable.  The coordinate
transformation decorrelates the components in the random field and makes the second obstacle
relatively easy to overcome. The distribution of each random variable can be estimated independently
from the transformed samples.  Finally, the estimated distribution can be sampled straightforwardly
to obtain new samples.  Sampling of a given distribution of a scalar random variable can be achieved
by using standard statistical techniques (e.g., transformation of a uniform random variable with the
inverse cumulative distribution function) or more advanced methods in the literature.
The detailed algorithm of the resampling procedure is presented in~\ref{sec:resample}.

The resampling procedure allows for the flexibility of using a larger number of samples for
propagating the posterior uncertainties in the Reynolds stresses than that used in the inference
procedure.  More importantly, when simultaneously propagating uncertainties in the input along with
model-form uncertainties, nested sampling or Sobol sampling~\cite{joe2008constructing} are needed
for the uncertainty propagation.  In this case it is important to be able to generate new samples
for the Reynolds stress discrepancy from the obtained distribution.  In the simulations presented in
this work, however, the process can be simplified by using the same posterior ensemble obtained from
the calibration for the uncertainty propagation in the prediction.  The resampling algorithm is
presented above for the completeness of the proposed framework.

\section{Numerical Results and Discussions}
\label{sec:veri}

In this section we demonstrate the capability of the proposed calibration--prediction method in two
scenarios as described in Section~\ref{sec:intro}. In the first scenario, the prediction flow and the calibration
flow are geometrically similar but have different Reynolds numbers. This scenario is relatively
straightforward and is examined on two cases, the flow over periodic hills and the flow in a square
duct. In the second scenario, the prediction case differs from the calibration case not only in
Reynolds number, but also in geometry.  Note that in all cases examined here, we assume that
observation data are only available in the calibration case and not in the prediction
case. Benchmark data in the prediction case are used only to assess the performance of the proposed
method.


\subsection{Prediction of Flow at Different Reynolds Numbers}
\label{sec:pred-re}

Two canonical flows of engineering relevance, the flow over periodic hills and the flow in a square
duct, are used to evaluate the performance of the proposed calibration-prediction method. Both flows
are challenging for the standard RANS models.  

\subsubsection{Flow over periodic hills}
\begin{figure}[htbp]
\centering
\includegraphics[width=0.75\textwidth]{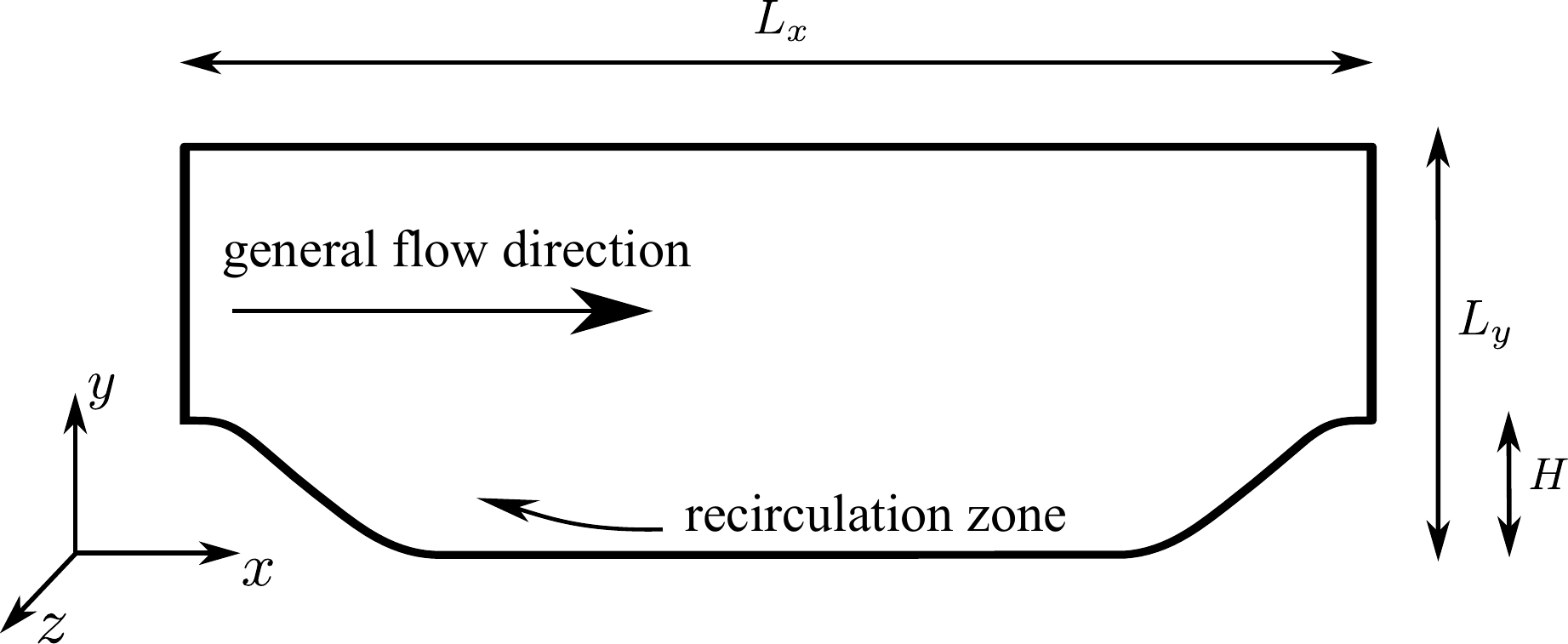}
\caption{Computational domain for the flow over periodic hills. The $x$, $y$ and $z$ coordinates
  are aligned with streamwise, wall-normal, and spanwise, respectively. The dimensions are
  normalized with $H$ with $L_x/H=9$, $L_y/H=3.036$, and $L_z/H=0.1$.}
\label{fig:domain_pehill}
\end{figure}

The flow over periodic hills is widely utilized to evaluate the performance of turbulence models due
to the comprehensive experimental and numerical benchmark data at a wide range of Reynolds
numbers~\cite{breuer2009flow}. The geometry of the computational domain is shown in
Fig.~\ref{fig:domain_pehill}. The Reynolds number $Re$ is based on the crest height $H$ and the bulk
velocity $U_b$ at the crest. Periodic boundary conditions are applied in the streamwise ($x$)
direction, and non-slip boundary conditions are applied at the walls. The spanwise ($z$) direction
is not considered since the mean flow is two-dimensional. The flow is calibrated on the flow at
$Re=2800$, and then predictions are made for the flow at $Re=10595$.  Predictions of the flow at
$Re=5600$ is also performed, and the results are qualitatively similar to those presented below for
the flow at $Re=10595$. Therefore, the results for the flow at $Re = 5600$ are omitted for brevity.

\begin{figure}[!htbp]
  \centering \hspace{2em}\includegraphics[width=0.5\textwidth]{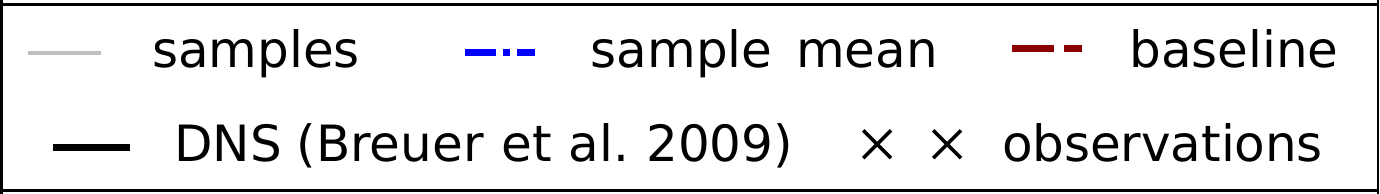}
  \subfloat[Prior ensemble of velocities]{\includegraphics[width=0.75\textwidth]{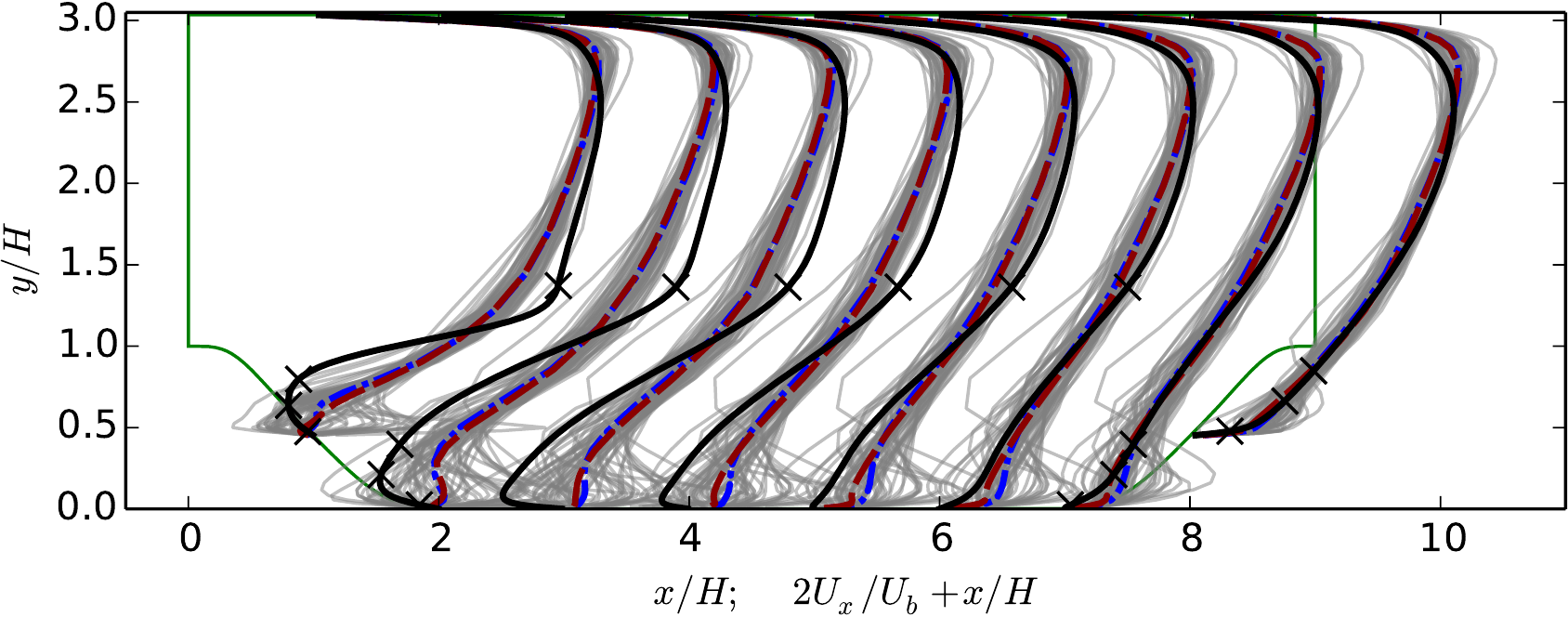}}\\
  \subfloat[Posterior ensemble
  velocities]{\includegraphics[width=0.75\textwidth]{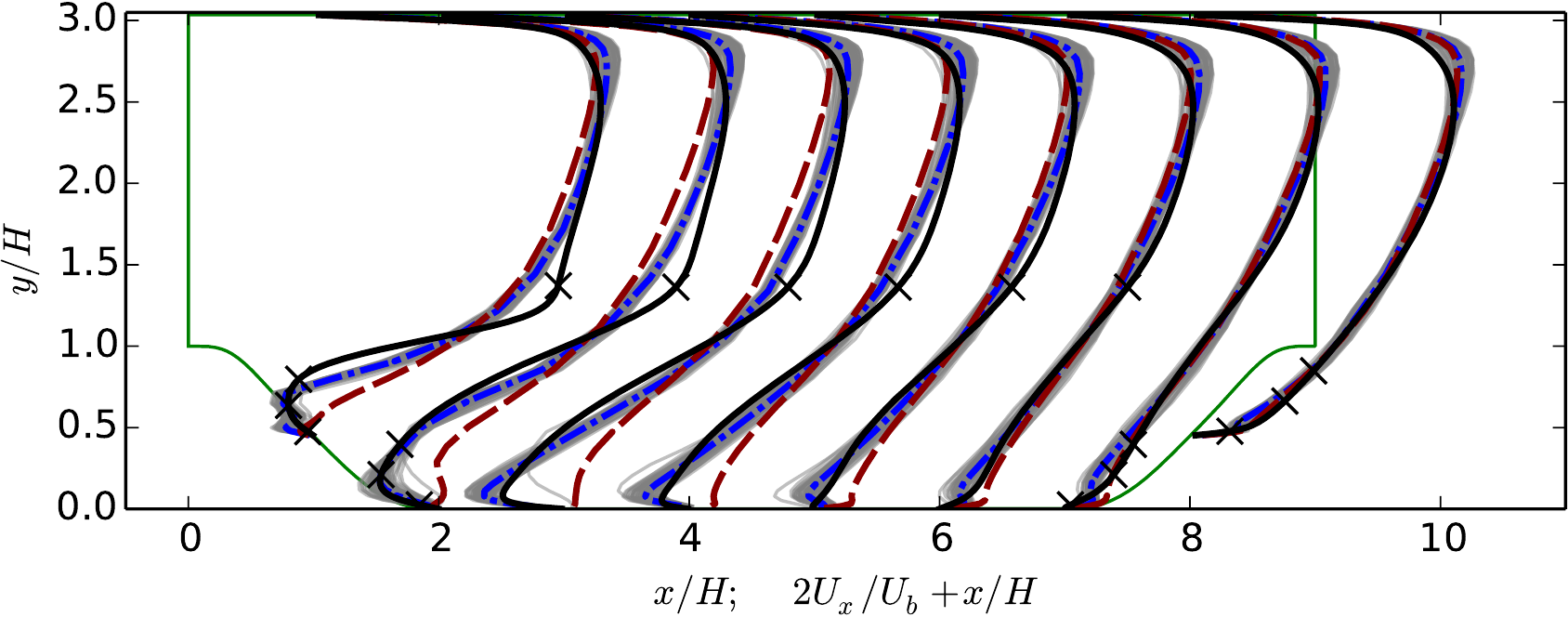}}
  \caption{Calibration of Reynolds stress discrepancies based on velocity observations in the flow
    over periodic hill at a low Reynolds number $Re=2800$.  This figure shows the (a) prior and (b)
    posterior ensembles of velocities at eight locations, $x/H=1,2, \cdots, 8$.  Black crosses
    ($\times$) denote locations where velocity observations are available.}
\label{fig:U_comp_pehill}
\end{figure}

The baseline RANS result of the flow at $Re=2800$ is calibrated with the procedure presented in
Section~\ref{sec:method-cali}. The prior and posterior ensembles of velocities are shown in
Fig.~\ref{fig:U_comp_pehill}. The prior ensemble of velocity profiles is scattered because of the
perturbations (uncertainties) introduced to the RANS-predicted Reynolds stresses. It can be seen
that the prior ensemble mean velocity is close to the baseline RANS results. Compared to the prior
ensemble mean velocity, the posterior mean velocity shows a much better agreement with the
benchmark, especially in the recirculation zone, where the magnitude of the reverse flow velocities
are significantly underestimated in the baseline RANS results.


\begin{figure}[!htbp]
\centering
\hspace{2em}\includegraphics[width=0.35\textwidth]{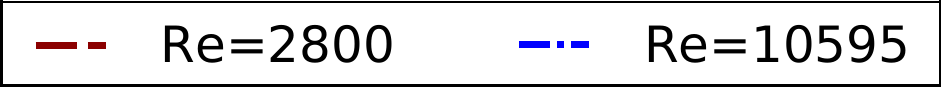}\\
\vspace{-0.8em}
\subfloat[Comparison of $\delta^{\xi}$]{\includegraphics[width=0.7\textwidth]{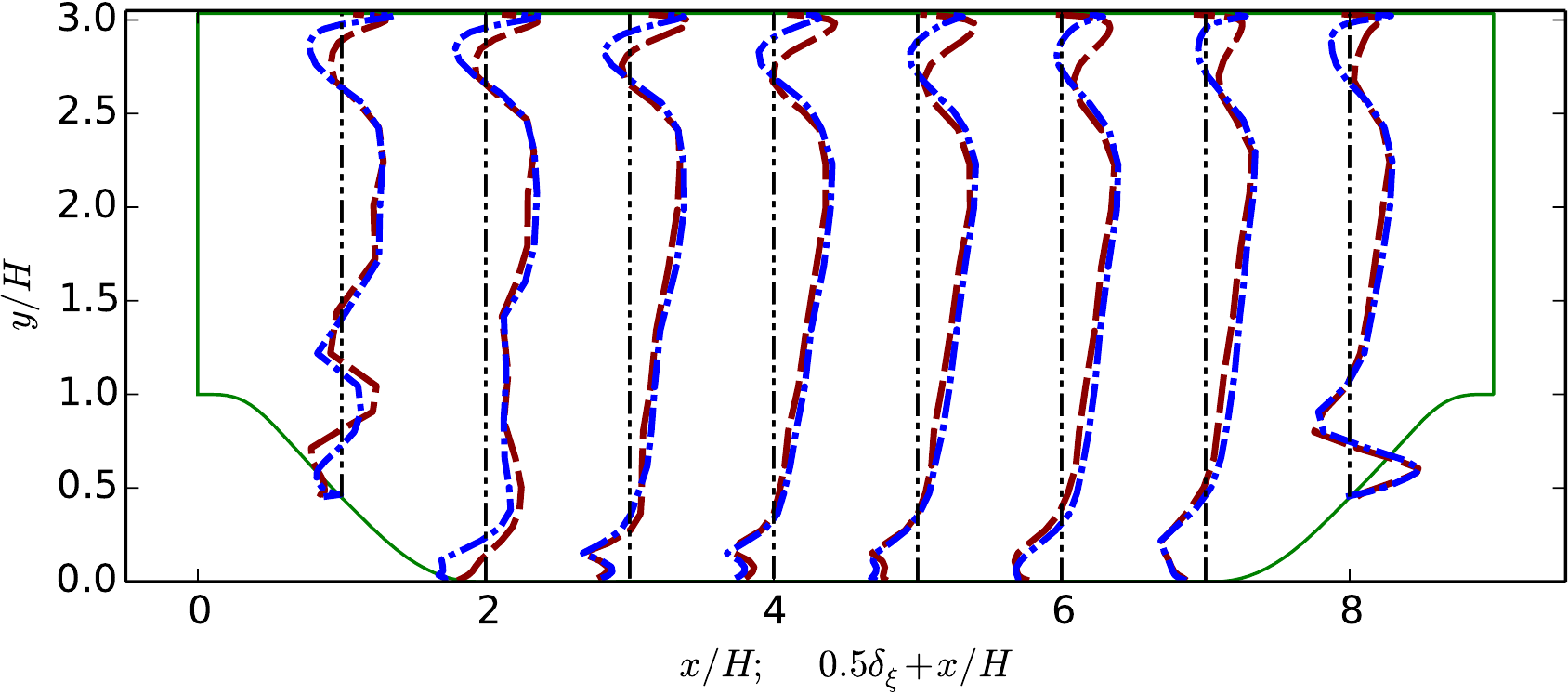}}\\
\subfloat[Comparison of $\delta^{\eta}$]{\includegraphics[width=0.7\textwidth]{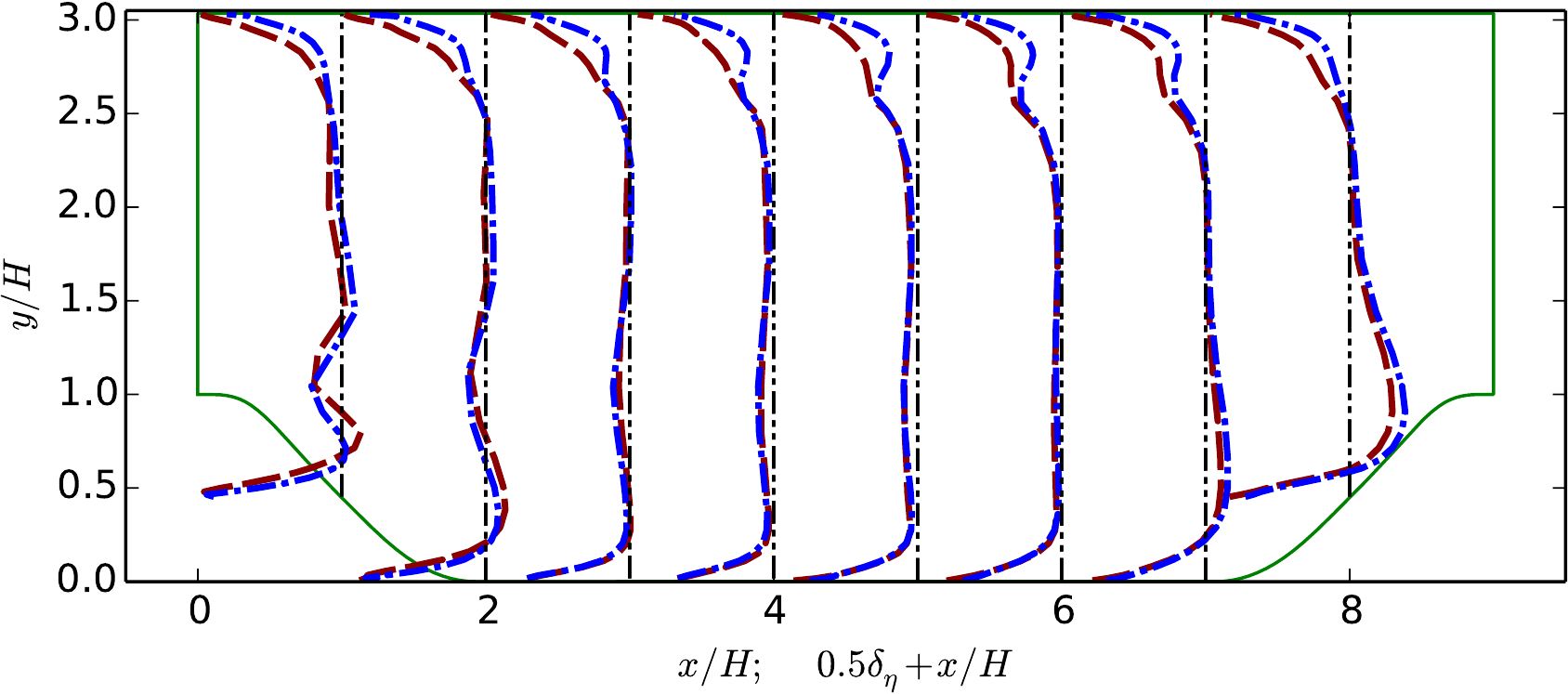}}\\
\subfloat[Comparison of $\delta^{k}$]{\includegraphics[width=0.7\textwidth]{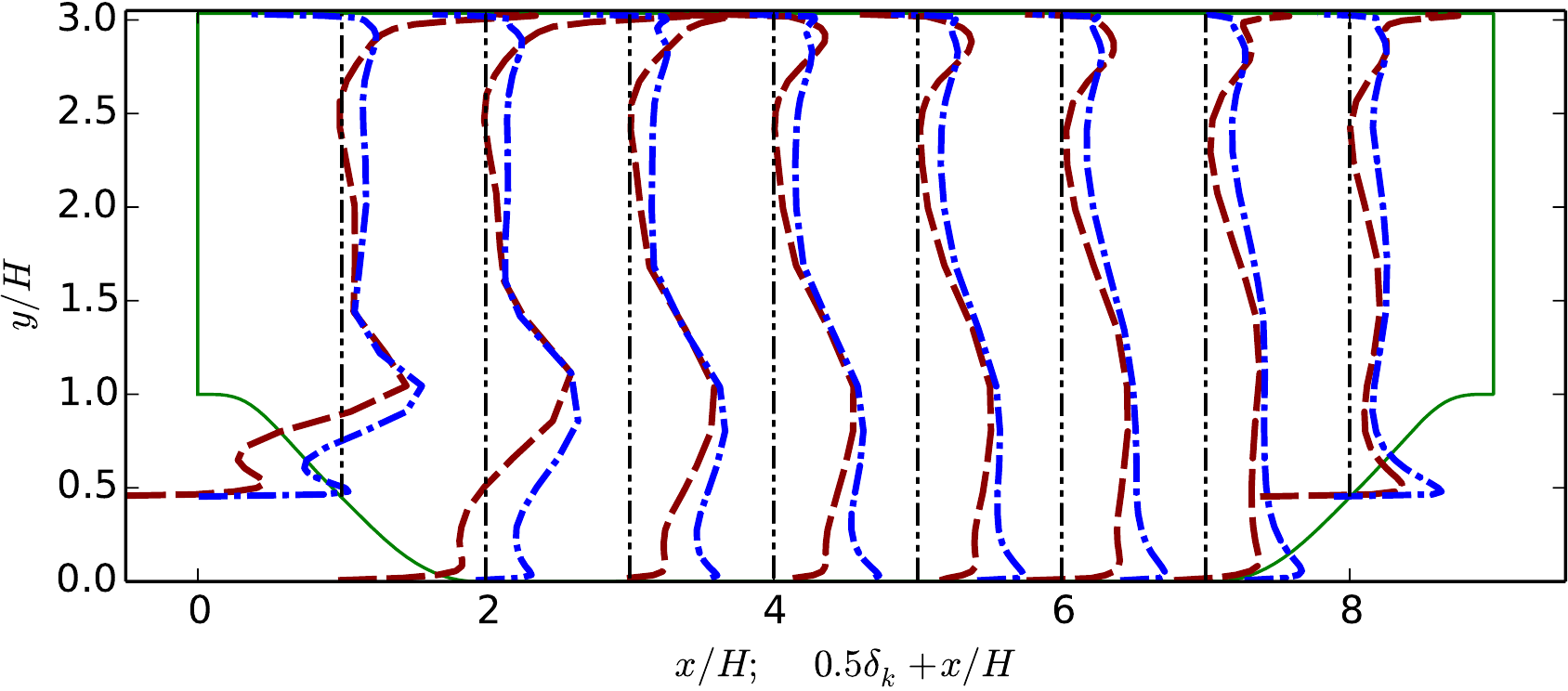}}
\caption{ The discrepancies of Reynolds stresses between baseline RANS and the benchmark data in the
  physically meaningful projections. This figure shows the discrepancies (a) $\delta^{\xi}$, (b)
  $\delta^{\eta}$, and (c) $\delta^k$ for the flows at $Re=2800$ and $Re=10595$, which are the
  calibration and prediction cases, respectively.}
\label{fig:para_diff}
\end{figure}

Figure~\ref{fig:para_diff} shows the Reynolds stresses discrepancies (i.e., the difference between
the baseline RANS prediction and the benchmark data) for the calibration case ($Re=2800$) and the
prediction case ($Re=10595$).  More precisely, the discrepancies $\delta^k$, $\delta^{\xi}$, and
$\delta^{\eta}$ in the physical projections are presented. As shown in Figs.~\ref{fig:para_diff}a
and~\ref{fig:para_diff}b, $\delta^{\xi}$ and $\delta^{\eta}$ are similar for the flows at the two
Reynolds numbers. It indicates that the discrepancies in the anisotropy of the Reynolds stresses are
not sensitive to a moderate change of Reynolds number from $2800$ to $10595$. This is because the
anisotropy is mainly influenced by the geometry of the domain and the coherent structures of the
flow. In contrast, the log-discrepancy $\delta^k$ in turbulent kinetic energy (TKE) is clearly
different for the two flows, especially in the recirculation region. This is related to a potential
deficiency of extrapolating the log-discrepancy of the turbulent kinetic energy.

It is not clear if the log-discrepancy of the TKE, its discrepancy, or a combination of both is the
best choice of quantity to extrapolate from the calibration to prediction case. As the
log-discrepancy is dimensionless while the discrepancy has a dimension of the TKE, the former seems
to be preferred. However, the hypothetical case shown in Table~\ref{tab:deltak} suggests
otherwise. Suppose in a particular region the true turbulence intensity, defined as the ratio of the
velocity fluctuations $u'$ to the reference velocity $U_b$, is 10\% in both the calibration and
prediction cases, while the RANS simulations give 0.1\% and 1\%, respectively.  The errors in the
RANS predictions are in fact rather consistent for the two cases in that the flow has a high
intensity turbulence while the predictions give low turbulence intensities. In other words, the
difference between $0.1\%$ and $1\%$ is not as significant as that between $1\%$ and $10\%$.  The
true log-discrepancy would be 2 and 1 for the calibration and prediction cases, respectively. Hence,
extrapolating the log-discrepancy to the prediction case would lead to a corrected turbulent
intensity of $100\%$ in the prediction case, which is drastically different from the truth, and such
a correction can destabilize the simulation.  In contrast, the absolute discrepancies are
approximately the same for both cases and is thus a more appropriate quantity to extrapolate.  On
the other hand, one can easily devise a scenario where extrapolating the log-discrepancy is more
reasonable. A more sophisticated scheme of extrapolation is needed and is the subject of future
work.  In view of the considerations above, the discrepancy of the turbulent kinetic energy is not
extrapolated from the calibration cases to the prediction cases in the simulations presented
below. In other words, the baseline RANS-modeled turbulent kinetic energy in the latter case is
not perturbed in the prediction.

\begin{table}[!htbp]
  \centering
  \begin{tabular}{l|cc}
    \hline
    & \textbf{calibration case} & \textbf{prediction case} \\
    \hline
    true turbulence intensity & 10\% & 10\% \\
    baseline RANS predicted intensity &  0.1\% &  1\% \\
    log-discrepancy of TKE  & 2 & 1 \\
    discrepancy of TKE & $\sim10^{-2} U_b^2$ &  $\sim10^{-2} U_b^2$ \\
    \hline
  \end{tabular}
  \caption{A hypothetical case to illustrate the potential deficiency of extrapolating log-discrepancy of the
    turbulent kinetic energy (TKE) from the calibration case to the prediction case. Turbulent intensities $u'/U_b$
    in both the calibration case and the prediction case are shown, where $u'$ denotes velocity
    fluctuation and $U_b$ is the reference bulk velocity.}
  \label{tab:deltak}
\end{table}

\begin{figure}[!htbp]
\centering
\hspace{2em}\subfloat{\includegraphics[width=0.7\textwidth]{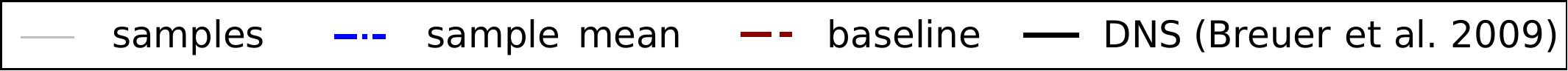}}\\
\vspace{-0.8em}
\subfloat{\includegraphics[width=0.8\textwidth]{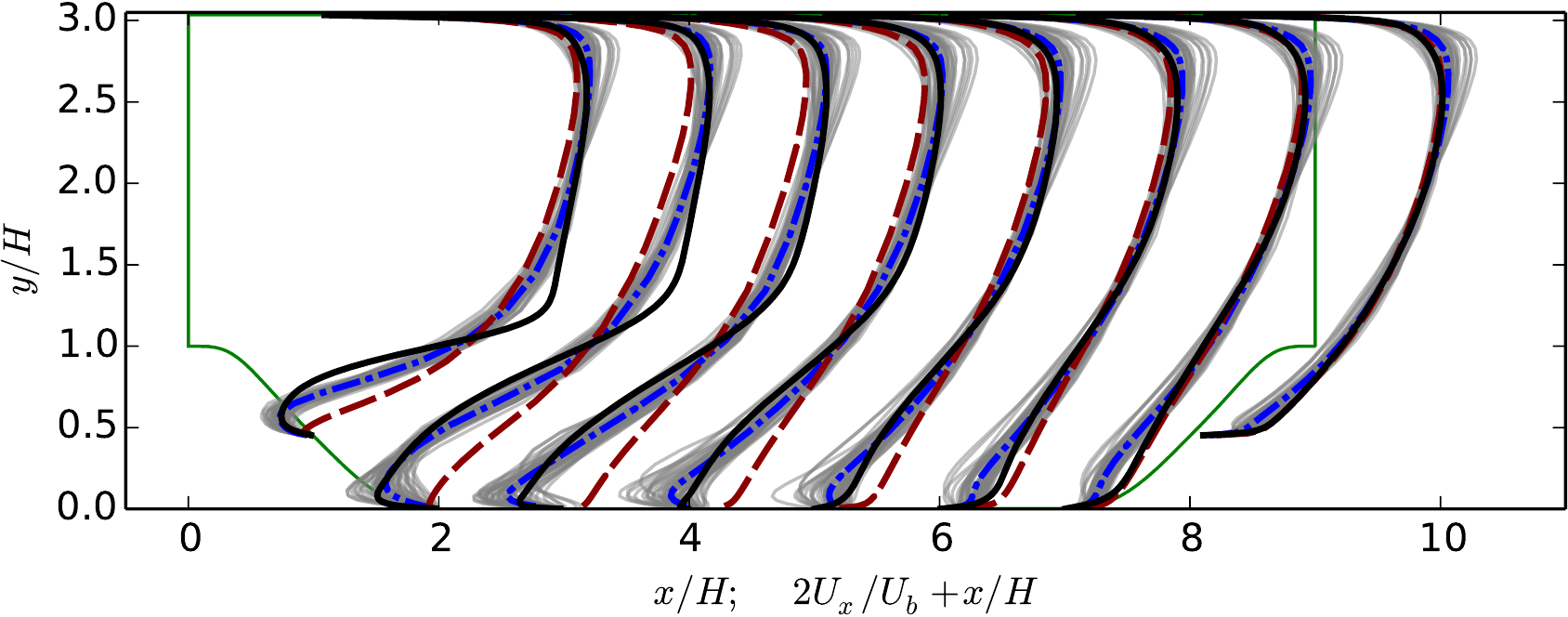}}
\caption{Ensemble of predicted velocity profiles of the flow over periodic hills of $Re=10595$ at
  eight streamwise cross-sections $x/H=1, 2, \cdots, 8$ compared with benchmark data and baseline
  results.
\label{fig:Ux_Re10595}
} 
\end{figure}

\begin{figure}[!htbp]
\centering
\includegraphics[width=0.75\textwidth]{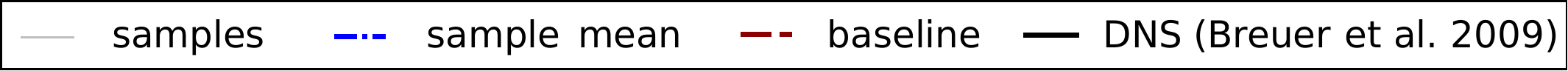}\\
\subfloat[Calibration case at $Re=2800$]{\includegraphics[width=0.45\textwidth]{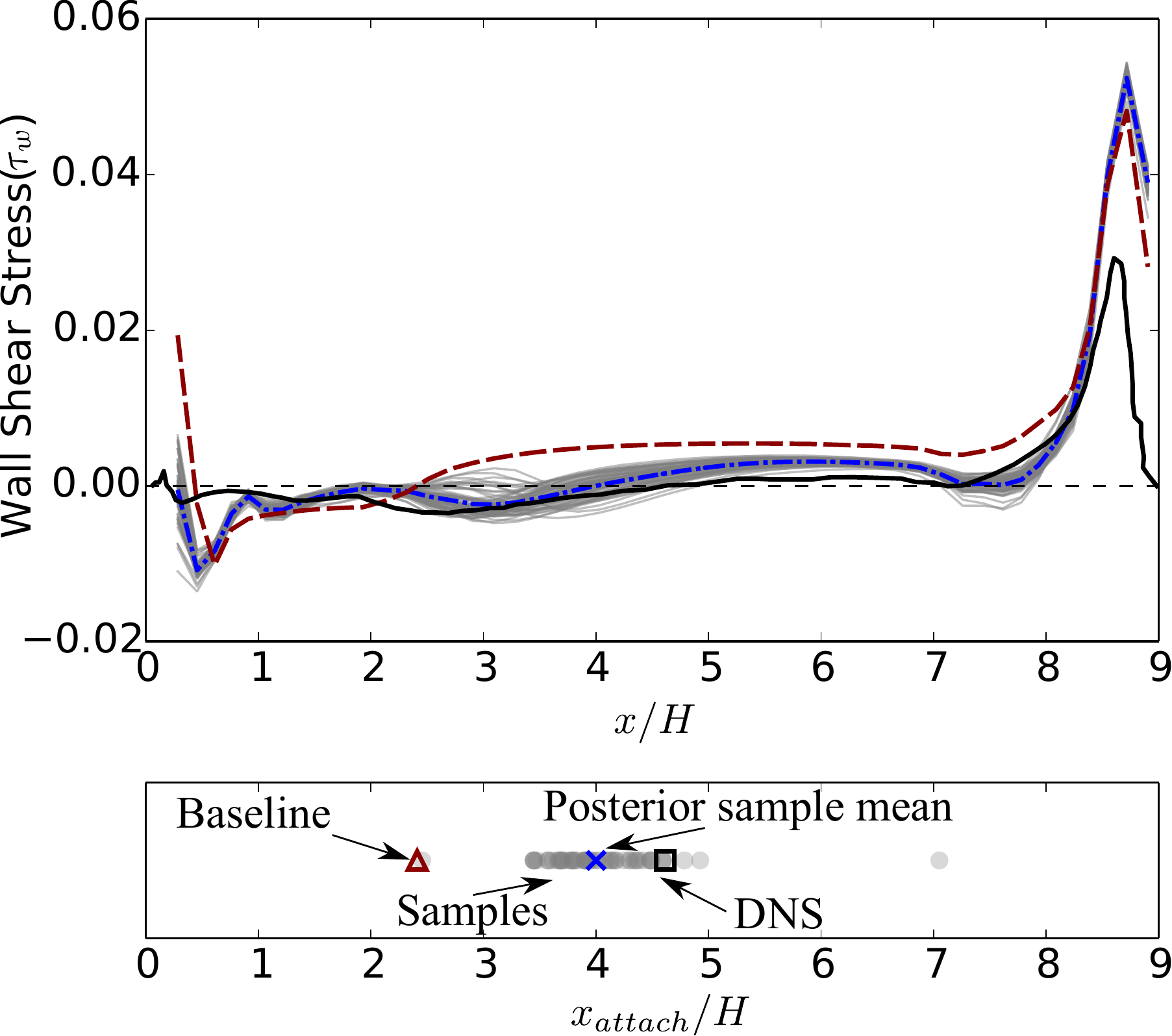}}\hspace{0.05em}
\subfloat[Prediction case at
$Re=10595$]{\includegraphics[width=0.45\textwidth]{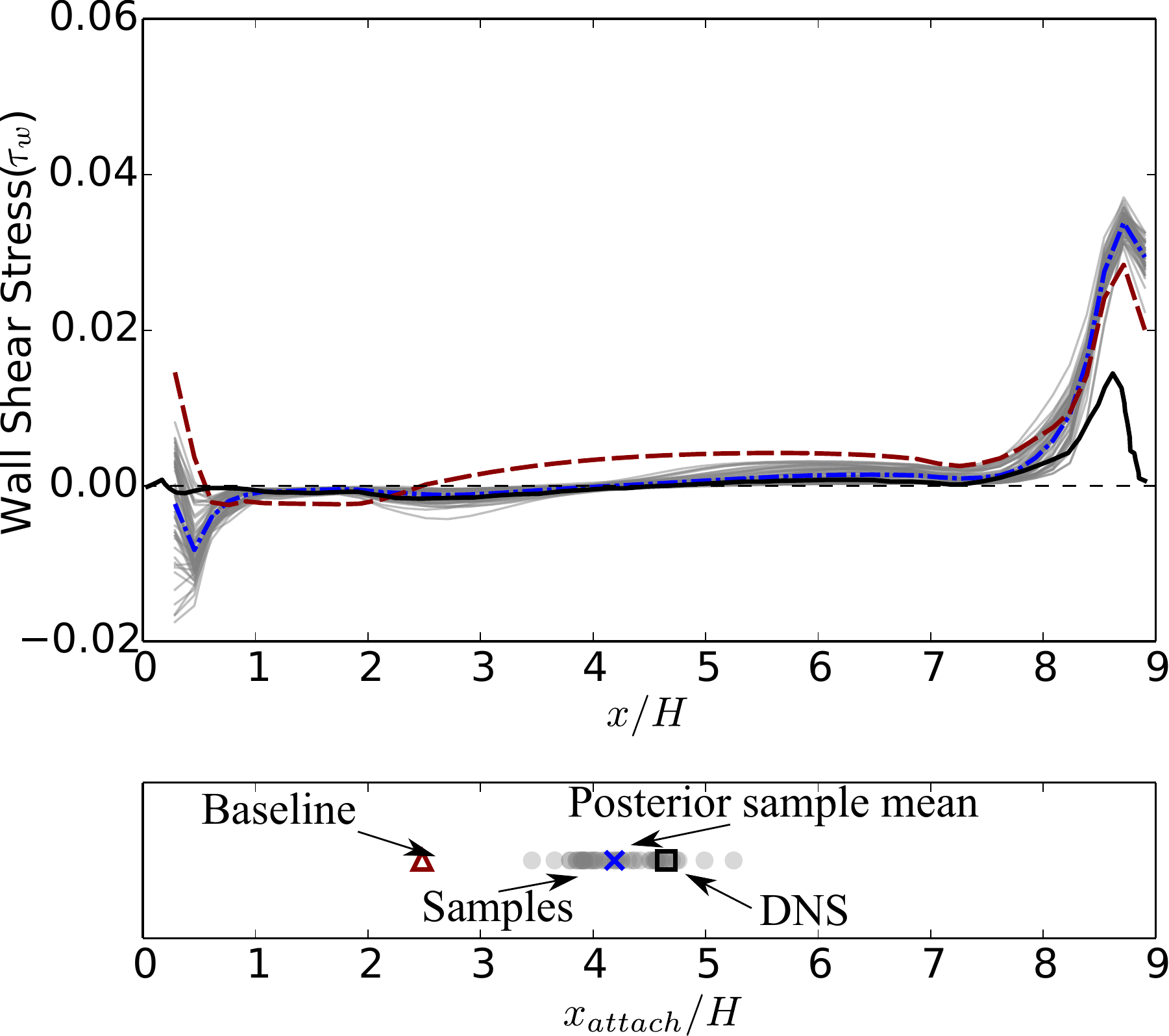}}
\caption{Posterior ensembles of shear stress $\tau_w$ on the bottom wall and reattachment points
  $x_{attach}$ for the flow over periodic hills, showing both (a) the calibration case at $Re=2800$
  and (b) the prediction case at $Re=10595$. The regions with negative wall shear stresses on the
  bottom wall are recirculation zones. The reattachment point is determined by the change of
  wall shear stress from negative to positive, which is the downstream end of the recirculation
  zone.}
\label{fig:bubble_Re10595}
\end{figure}

The flow over periodic hills at $Re=10595$ is predicted based on the calibrated Reynolds stresses
discrepancies, and the predictions of QoIs including the velocities, wall shear stresses, and
reattachment location are compared to the benchmark data~\cite{breuer2009flow}. The posterior
ensemble of velocities are shown in Fig.~\ref{fig:Ux_Re10595}. It can be seen that the baseline RANS
simulation underestimates the velocity magnitude in the recirculation zone, or equivalently the
strength of the recirculation. Compared to the baseline RANS results, the posterior mean profiles of
the velocities have significantly improved agreement with the benchmark data, especially in the
recirculation zone. Admittedly, differences still exist between the posterior ensemble mean and the
benchmark data. For example, the velocities near the flow reattachment between $x/H=4$ and $7$ are
over-corrected. Such differences are expected because all samples and the mean of the posterior
ensemble in the Kalman ensemble inference lie in the space spanned by the prior ensemble, but the
truth may reside outside this space.  More detailed discussions on this issue can be found
in~\cite{xiao-mfu}.  Figure~\ref{fig:bubble_Re10595} shows the bottom wall shear stress $\tau_w$ and
the reattachment point $x_{attach}$ of the flows in both the calibration case ($Re=2800$) and the
prediction case ($Re=10595$). The recirculation zones are indicated by the range in which the wall
shear stresses are negative. It can be seen that baseline RANS simulations underpredict the sizes of
the recirculation zones in both the calibration and the prediction cases. Even without incorporating
direct observation data, the predicted recirculation zone size is significantly improved. Such
improvement is also confirmed by the posterior ensemble mean of reattachment point, which is much
closer to the benchmark data compared to the baseline RANS result. This improvement is particularly
notable considering the fact that velocity observations are not used for the inference in the
prediction case ($Re=10595$).

The main difference between the prediction and the benchmark data is near the crest of the hill,
i.e., on the leeward side from $x/H=0$ to $x/H=1$ and on the windward side from $x/H=8$ to $x/H=9$.
This is attributed to the lack of observation data at these locations in the calibration flow
($Re=2800$). The wall shear stress $\tau_w$ samples are even more scattered at these locations
compared to those of the calibration case. A possible explanation is that the Reynolds number in the
prediction case is higher than that in the calibration case, and thus the length
scale of prediction case is smaller. Consequently, spatial correlations of velocities are small between the 
region near the crest
and the locations where observation data are available (e.g., in the free shear region and the
recirculation zone). The weak correlation leads to larger scattering of predicted wall shear
stresses $\tau_w$.

\subsubsection{Flow in Square Duct}

\begin{figure}[htbp]
\centering
\includegraphics[width=0.8\textwidth]{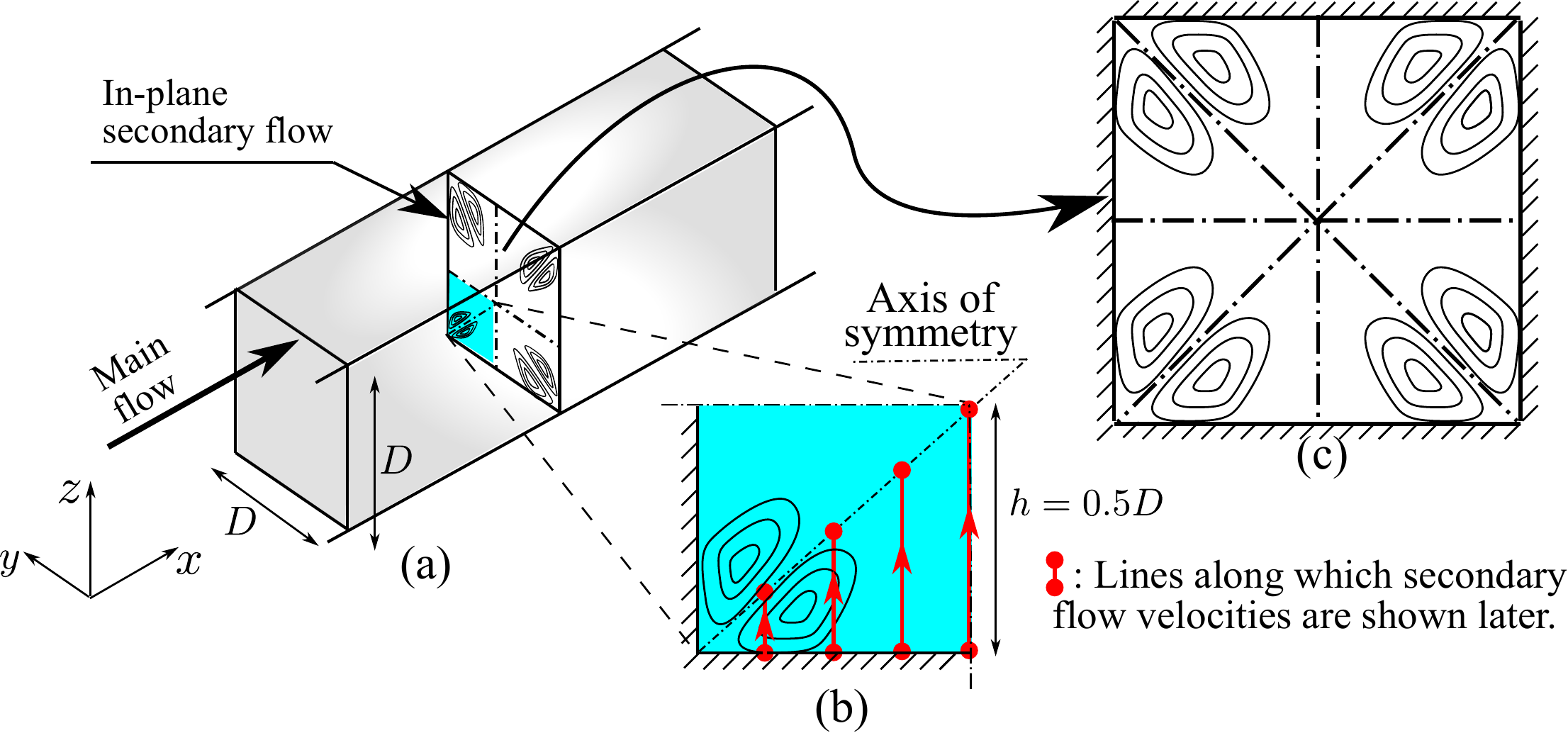}
\caption{Domain shape for the flow in a square duct. The $x$ coordinate represents the streamwise
  direction. Secondary flows induced by Reynolds stress imbalance exist in the $y$--$z$
  plane. Panel (b) shows that the computational domain covers a quarter of the cross-section of the
  physical domain. This is due to the symmetry of the mean flow in both $y$ and $z$ directions as
  shown in panel (c).}
\label{fig:domain_duct}
\end{figure}

The fully developed turbulent flow in a square duct is a widely known case for which RANS models
fail to predict the secondary flow induced by Reynolds stress imbalances~\cite{huser1993direct}.  A
schematic is presented in Fig.~\ref{fig:domain_duct} to show the physical domain, major features of
the flow, and the dimensions of the computational domain. Since the flow is fully developed in the
streamwise direction, a two-dimensional simulation is performed.  The computational domain only
covers a quarter of the cross-section as shown in Fig.~\ref{fig:domain_duct}b based on the symmetry
of the computational domain along $y$ and $z$ directions. All lengths are normalized by the height
of the computational domain $h=0.5D$, where $D$ is the height of the duct. The Reynolds number $Re$
is based on duct height $D$ and bulk velocity $U_b$.  The Reynolds stress discrepancies are
calibrated on the flow at $Re=1 \times 10^4$, and predictions are made for flows at $Re=8.3 \times
10^4$ and $2.5 \times 10^5$.

\begin{figure}[!htbp]
\centering
\includegraphics[width=0.5\textwidth]{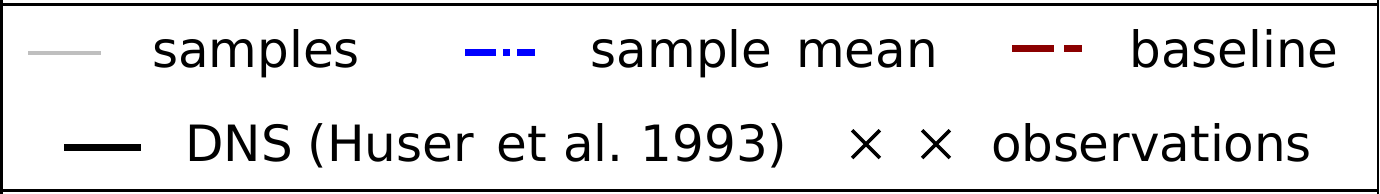}\\
\subfloat[Prior ensemble]{\includegraphics[width=0.45\textwidth]{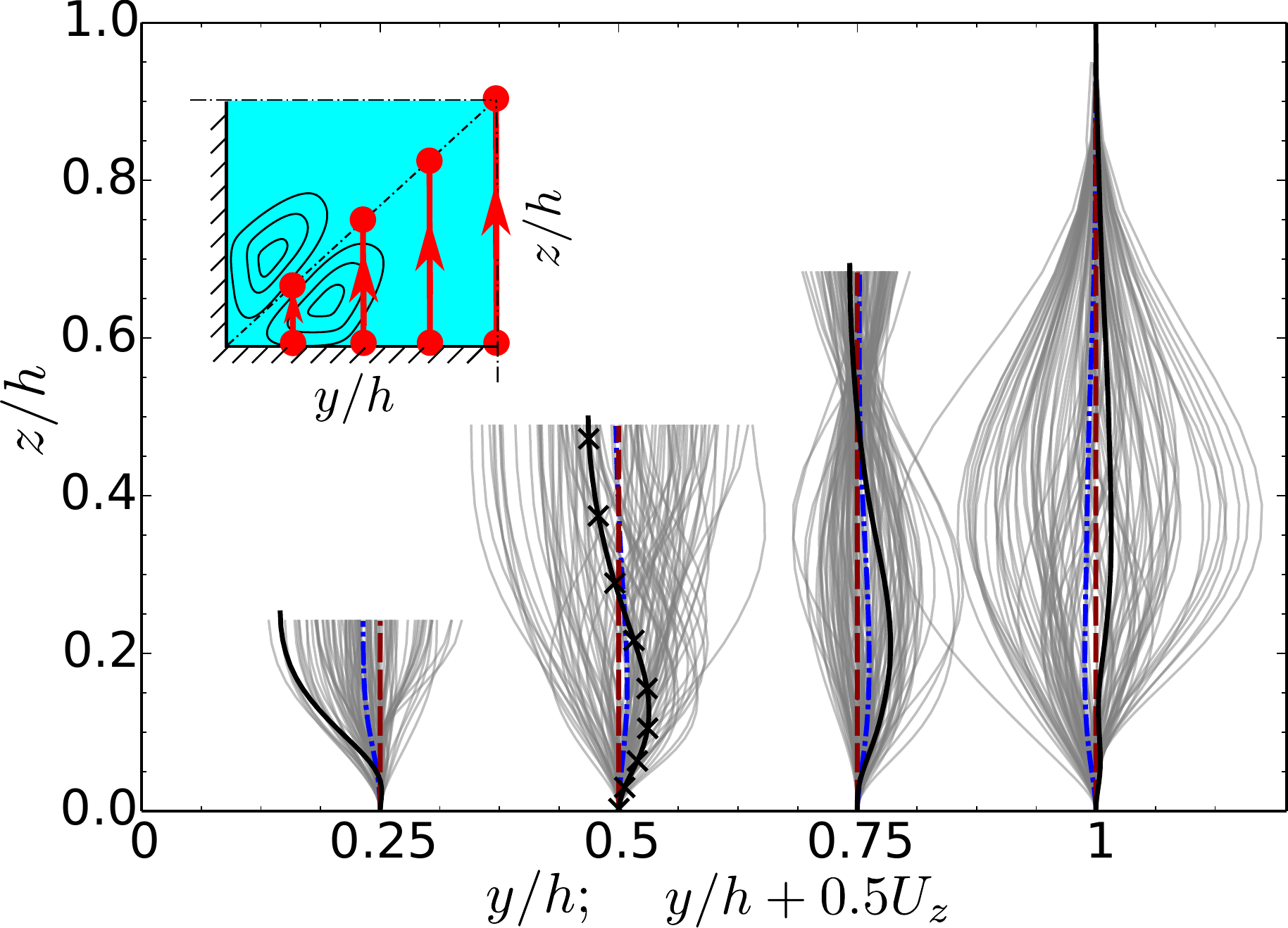}}\hspace{0.05em}
\subfloat[Posterior ensemble]{\includegraphics[width=0.45\textwidth]{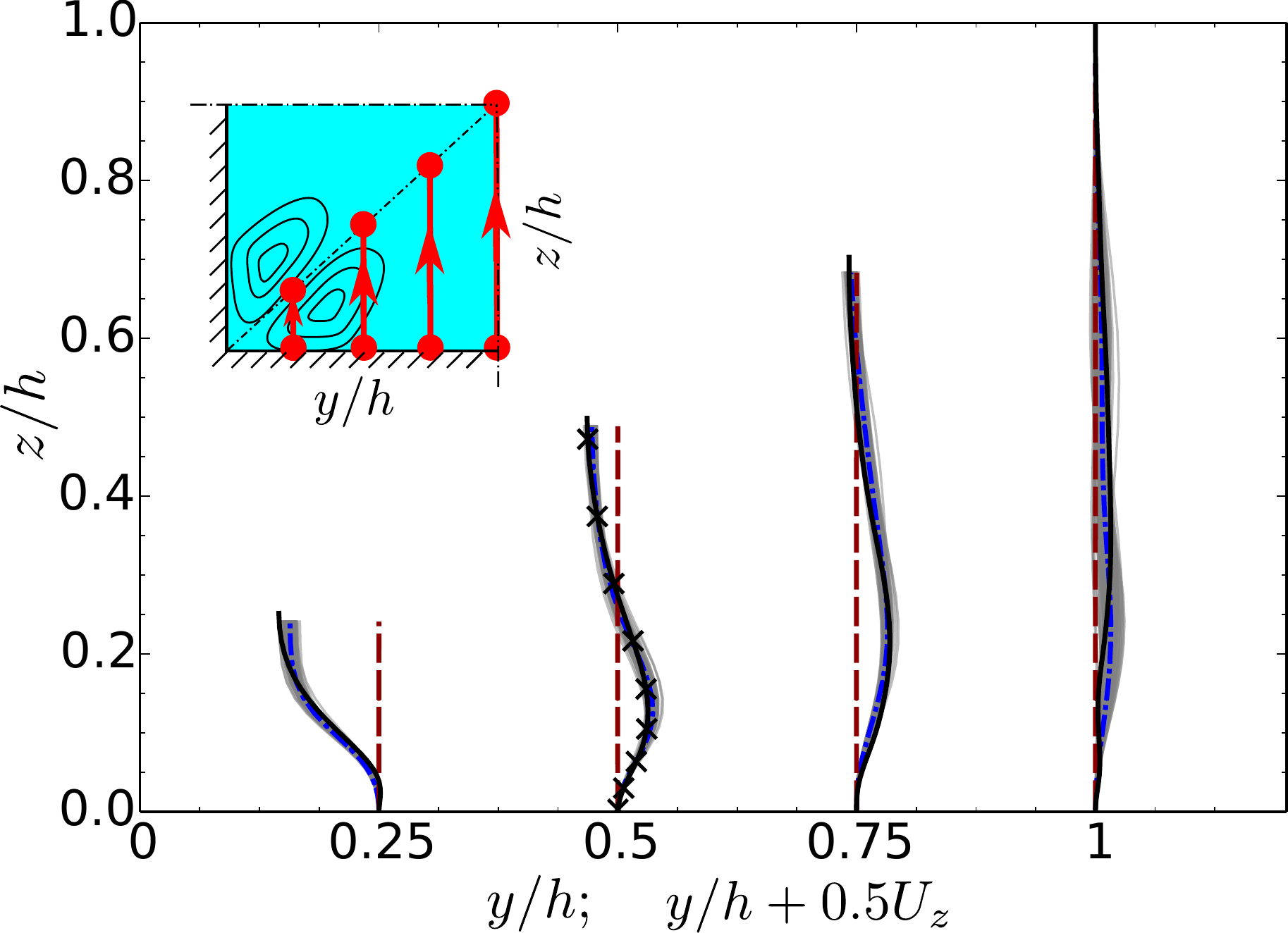}}
\caption{(a) Prior ensemble and (b) posterior ensemble of velocity $U_z$ profiles at four locations
  $y/h=0.25, 0.5, 0.75$ and $1$.  The sample profiles of prior are scaled by a factor of 0.3 for
  clarity.}
\label{fig:U_duct}
\end{figure}

As in the periodic hill cases presented above, we only consider uncertainties in parameters $\xi$
and $\eta$, which represent the anisotropy (i.e., shape) of the Reynolds stress tensor.  This is
because the secondary flow is primarily induced by the anisotropy of Reynolds
stresses. Figure~\ref{fig:U_duct} shows the secondary velocities $U_z$ along four lines. It can be
seen from Fig.~\ref{fig:U_duct}a that the prior ensemble has a large range of scattering, which is
attributed to the sensitivity of the secondary flow to the turbulence anisotropy. Compared to the
prior ensemble mean, the posterior ensemble mean is much closer to the benchmark data, which
suggests that the discrepancies $\delta^{\xi}$ and $\delta^{\eta}$ are successfully calibrated.
With the calibrated discrepancies, predictions are made for the two flows at higher Reynolds
numbers.

\begin{figure}[!htbp]
  \centering
  \hspace{1em}\includegraphics[width=0.75\textwidth]{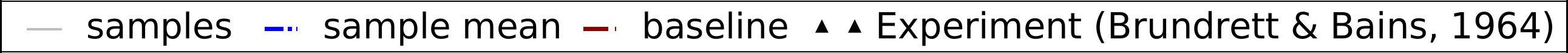}\\
  \vspace{-0.5em} \subfloat[Along vertical axis of
  symmetry]{\includegraphics[width=0.45\textwidth]{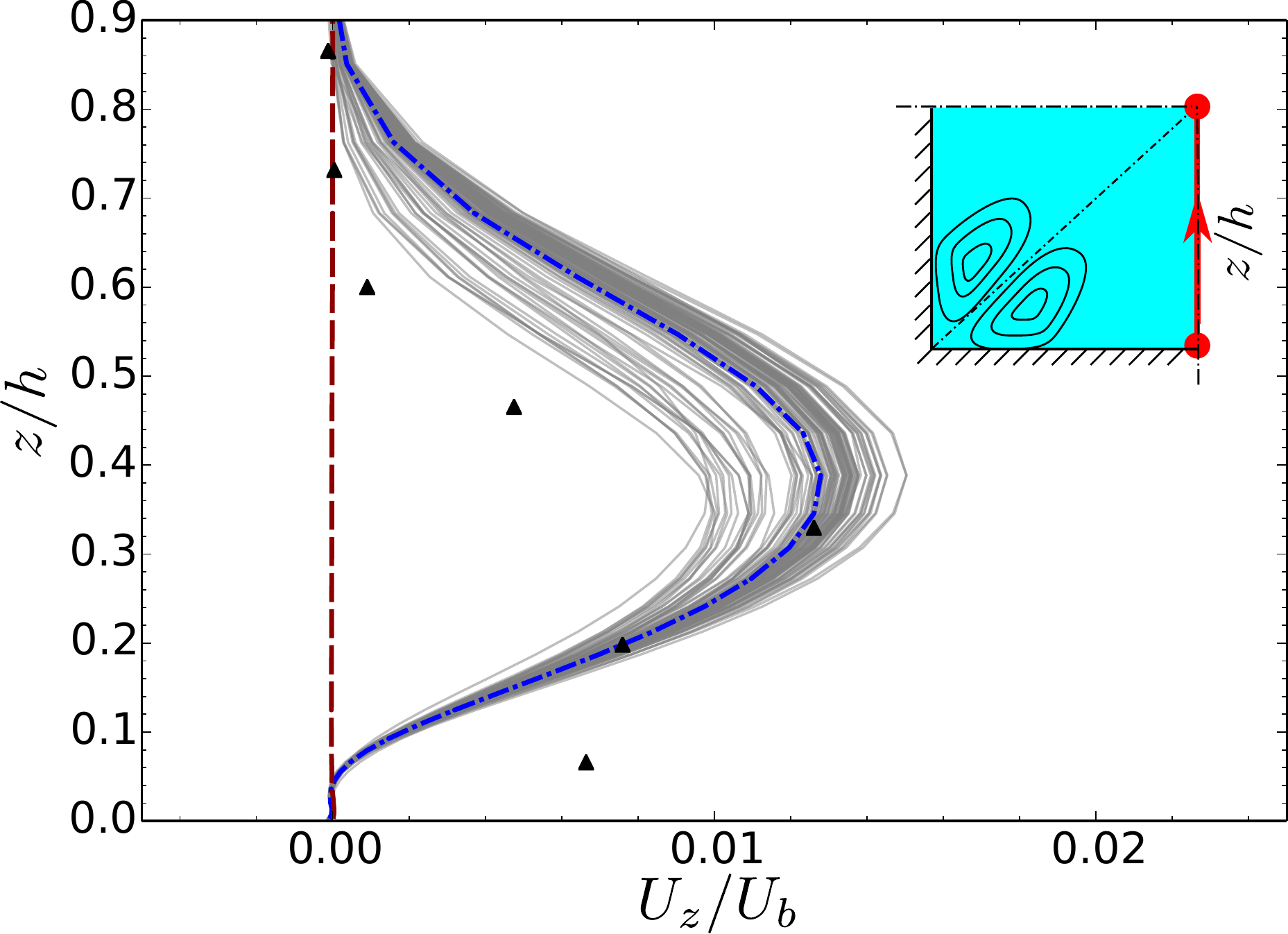}}\hspace{0.5em}
  \subfloat[Along diagonal]{\includegraphics[width=0.45\textwidth]{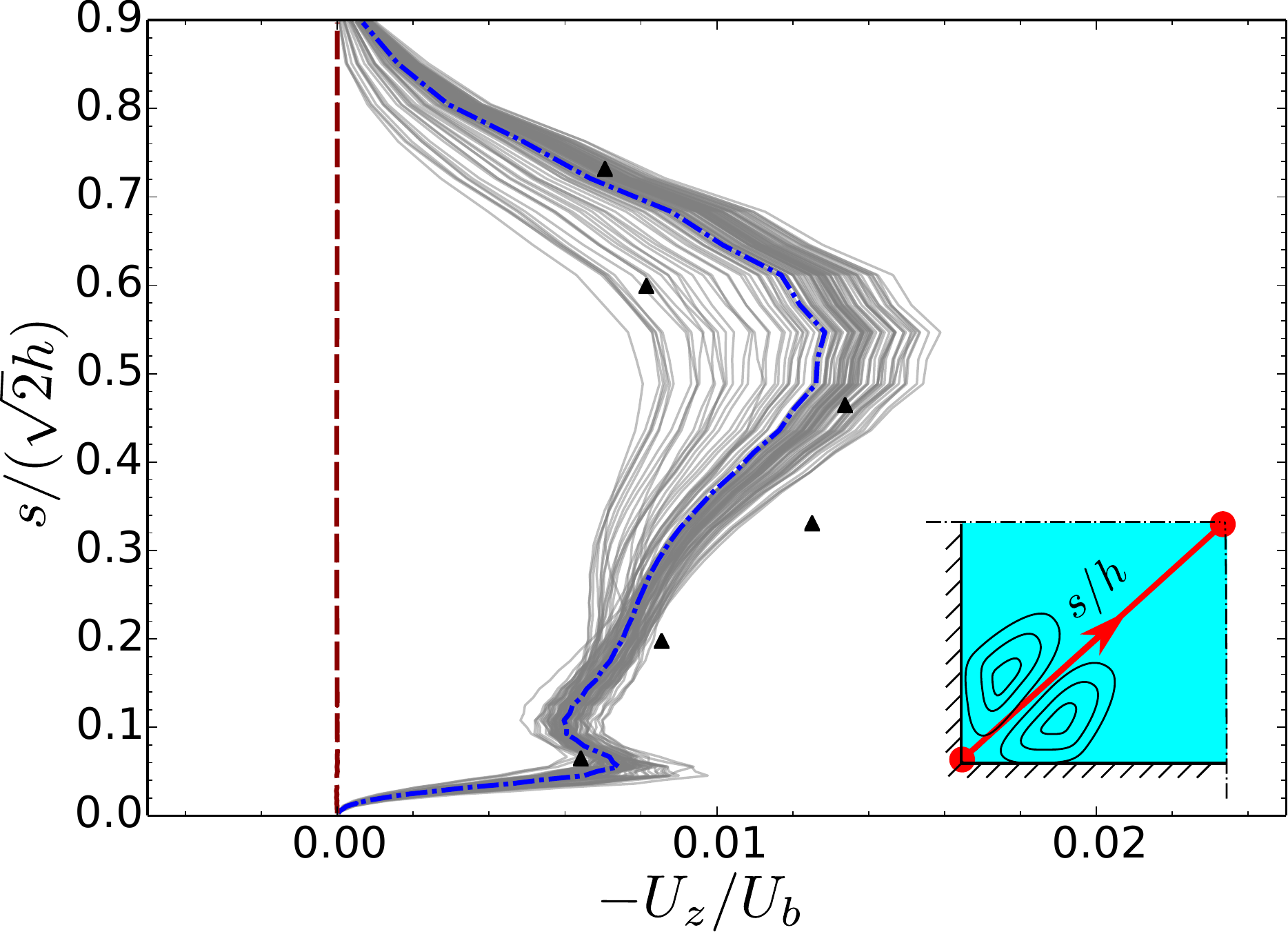}}\\
  \caption{The secondary flow velocity $U_y$ in the prediction case (flow at $Re=8.3 \times 10^4$,
    where no observation data are used). Velocity along (a) the vertical axis of symmetry and (b)
    the diagonal are shown. The experimental data are marked as $\blacktriangle$.}
\label{fig:Uy_Re3860}
\end{figure}

\begin{figure}[!htbp]
  \centering
  \hspace{1em}\includegraphics[width=0.75\textwidth]{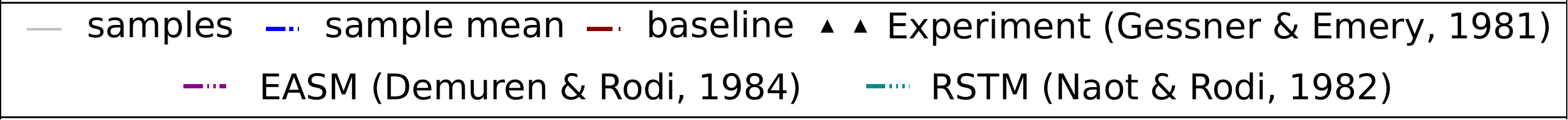}\\
  \vspace{-0.5em} \subfloat[Along vertical axis of symmetry
  ]{\includegraphics[width=0.45\textwidth]{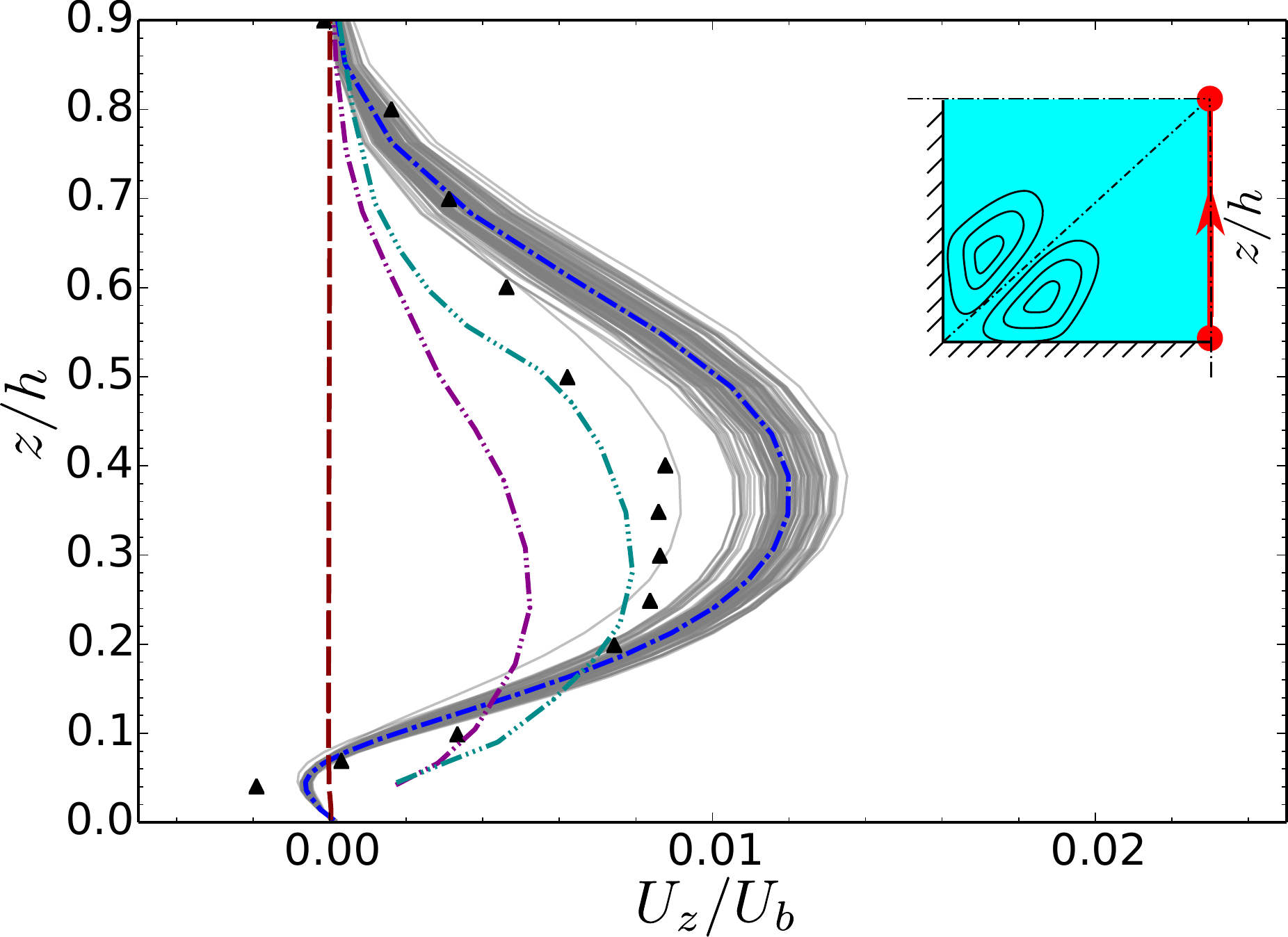}}\hspace{0.5em}
  \subfloat[Along diagonal]{\includegraphics[width=0.45\textwidth]{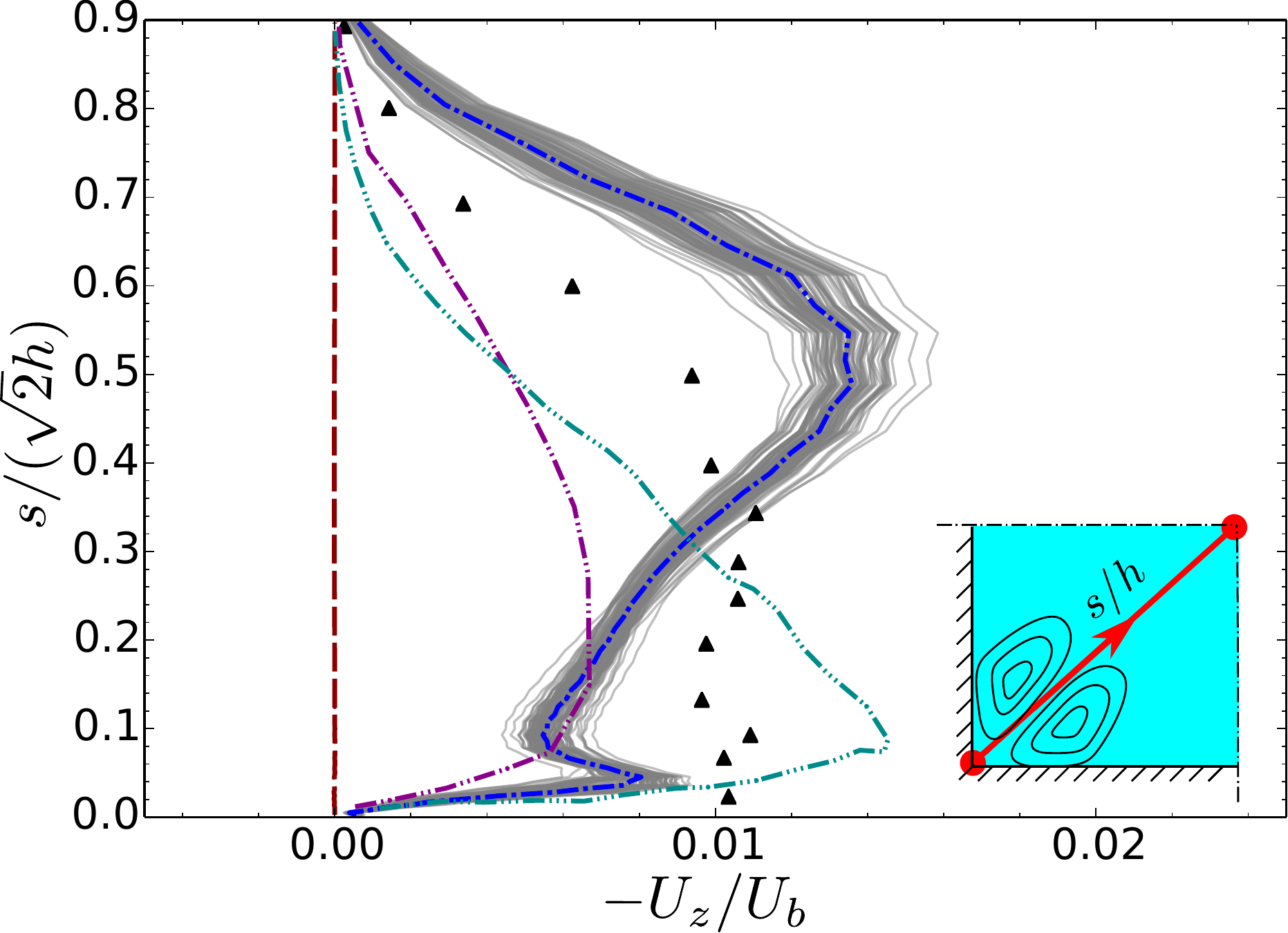}}\\
  \caption{The comparison of predicted secondary velocity $U_z$ at Reynolds number
    $Re=2.5\times10^5$ with experimental data (denoted as $\blacktriangle$) and with predictions
    from advanced turbulence models including an explicit algebraic Reynolds stress model (EASM) and
    a Reynolds stress transport model (RSTM). Comparisons are shown (a) along the vertical axis of
    symmetry and (b) along the diagonal.  Note that  the prediction does not utilize observation data from the flow at
    $Re=2.5\times10^5$.
\label{fig:Uy_Re10550}
}
\end{figure}

Figure~\ref{fig:Uy_Re3860}a shows the secondary velocity at $Re=8.3 \times 10^4$ along the vertical
axis of symmetry along $z=h$ (position shown in inset) with comparison to available experimental
data of Brundrett and Bains~\cite{brundrett64}.  The baseline RANS result deviates from the
experimental data because RANS models with isotropic eddy viscosity are not able to capture the
stress anisotropy induced secondary flows. In contrast, the present method predicts an in-plane
velocity magnitude comparable to the experimental data and captures its trend of spatial variation.
Similarly, the prediction for the flow at $Re=2.5 \times 10^5$ along the same line is presented in
Fig.~\ref{fig:Uy_Re10550}a, which also shows a good agreement with the benchmark data, even though
the Reynolds number in this case is more than an order of magnitude larger than that in the
calibration case.

Figure~\ref{fig:Uy_Re3860}b shows the secondary velocity at $Re=8.3 \times 10^4$ along the diagonal
of the channel with comparison to experimental data~\cite{gessner81}.  It can be seen that the
prediction is significantly improved compared to the baseline RANS result. The prediction for the
flow at $Re=2.5 \times 10^5$ as presented in Fig.~\ref{fig:Uy_Re10550}b also demonstrates a
satisfactory agreement with the benchmark data, although the location of the velocity peak is
slightly different from the benchmark results. A possible explanation for this difference is that
the assumption of similar Reynolds stress discrepancies becomes less accurate with the departure of
Reynolds number from that of the calibration flow. Despite this discrepancy, the
  prediction presented in Fig.~\ref{fig:Uy_Re10550}a and~\ref{fig:Uy_Re10550}b are still comparable
  to or better than those obtained from RANS solvers utilizing advanced turbulence models, including
  an explicit algebraic Reynolds stress model (EASM)~\cite{demuren1984} and a Reynolds stress
  transport model (RSTM)~\cite{naot1982}. It can be seen from Fig.~\ref{fig:Uy_Re10550}a that the
  secondary velocity $U_z$ of experimental data becomes negative between $z/h=0$ and $z/h=0.1$, a
  feature that is not captured by the explicit algebraic Reynolds stress model~\cite{demuren1984} or
  the Reynolds stress transport model~\cite{naot1982}.  Figure~\ref{fig:Uy_Re10550}b shows that the
  velocity drop between $z/h=0.1$ and $z/h=0.2$ as in the experimental data is only captured by the
  present method, although the both Reynolds stress models capture the magnitude and general trend
  of the experimental data well. It is worth noting that by combining a RANS solver with a simple
eddy viscosity model and indirect measurement data from a related flow (at a much lower Reynolds
number), the present framework is able to achieve better predictions of in-plane velocities than
advanced RANS models such as algebraic Reynolds stress model~\cite{demuren1984} and Reynolds stress
transport models~\cite{naot1982}. More assessment of advanced RANS models and comparison of their
simulation results to experimental data can be found in ref.~\cite{demuren1984}. This comparison
here highlights the merits of data in reducing model-form uncertainties in turbulence modeling.

\begin{figure}[!htbp]
  \centering 
  \subfloat[Calibration ($Re= 1 \times 10^4$)]{\includegraphics[width=0.32\textwidth]{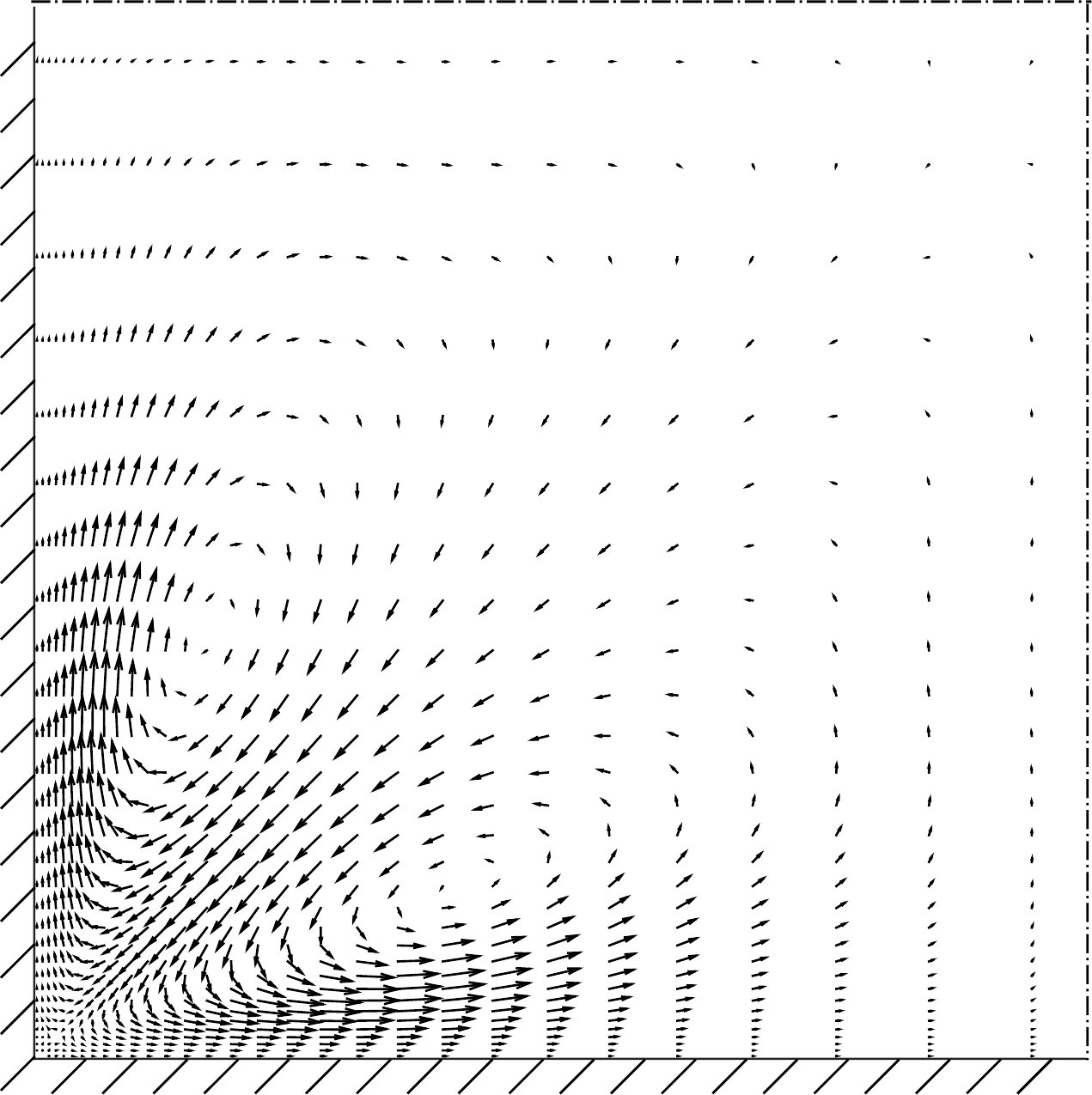}}\hspace{0.5em}
  \subfloat[Prediction ($Re=8.3 \times 10^4$)]{\includegraphics[width=0.32\textwidth]{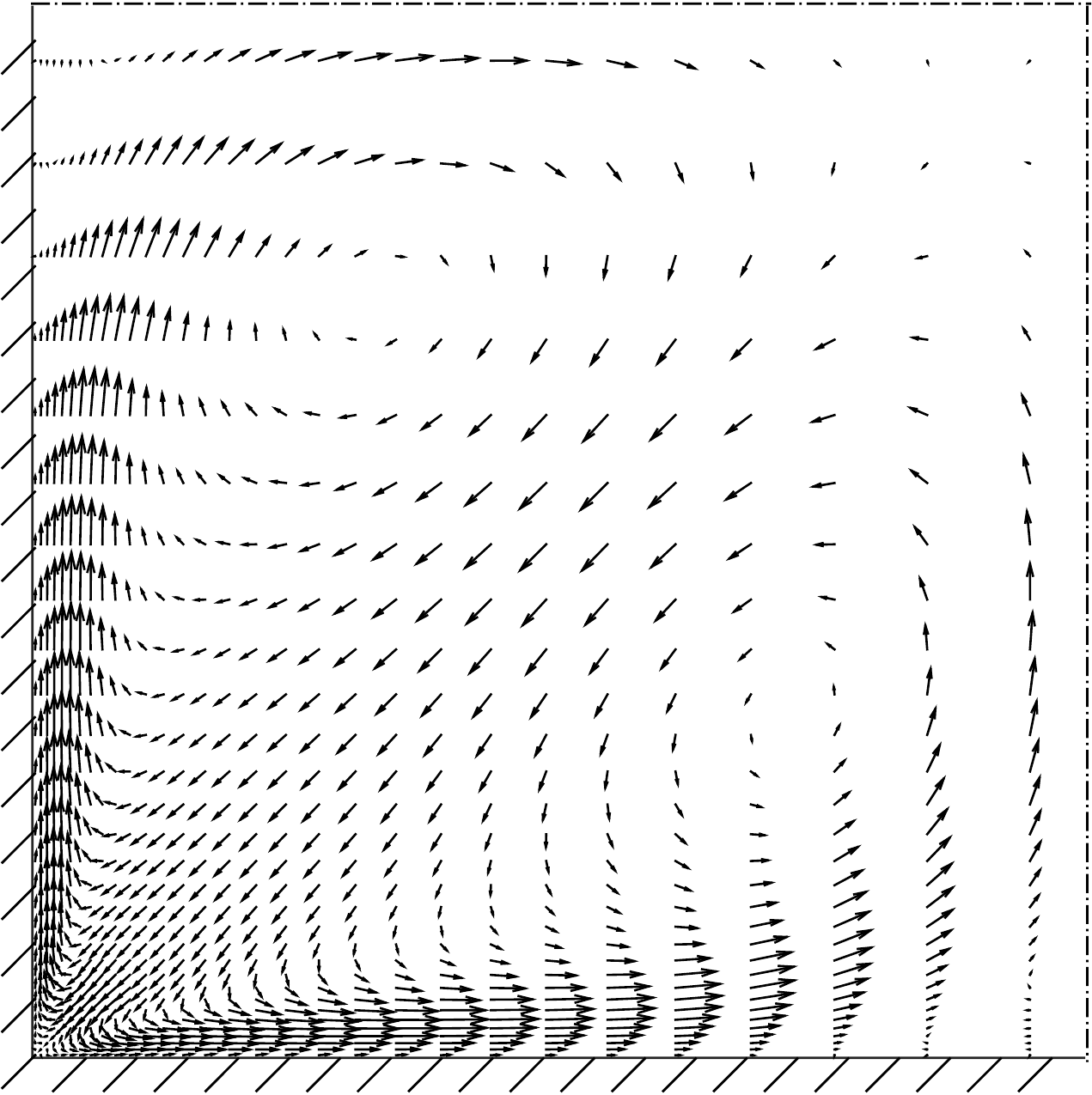}}\\
  \subfloat[Prediction ($Re=2.5 \times 10^5$)]{\includegraphics[width=0.32\textwidth]{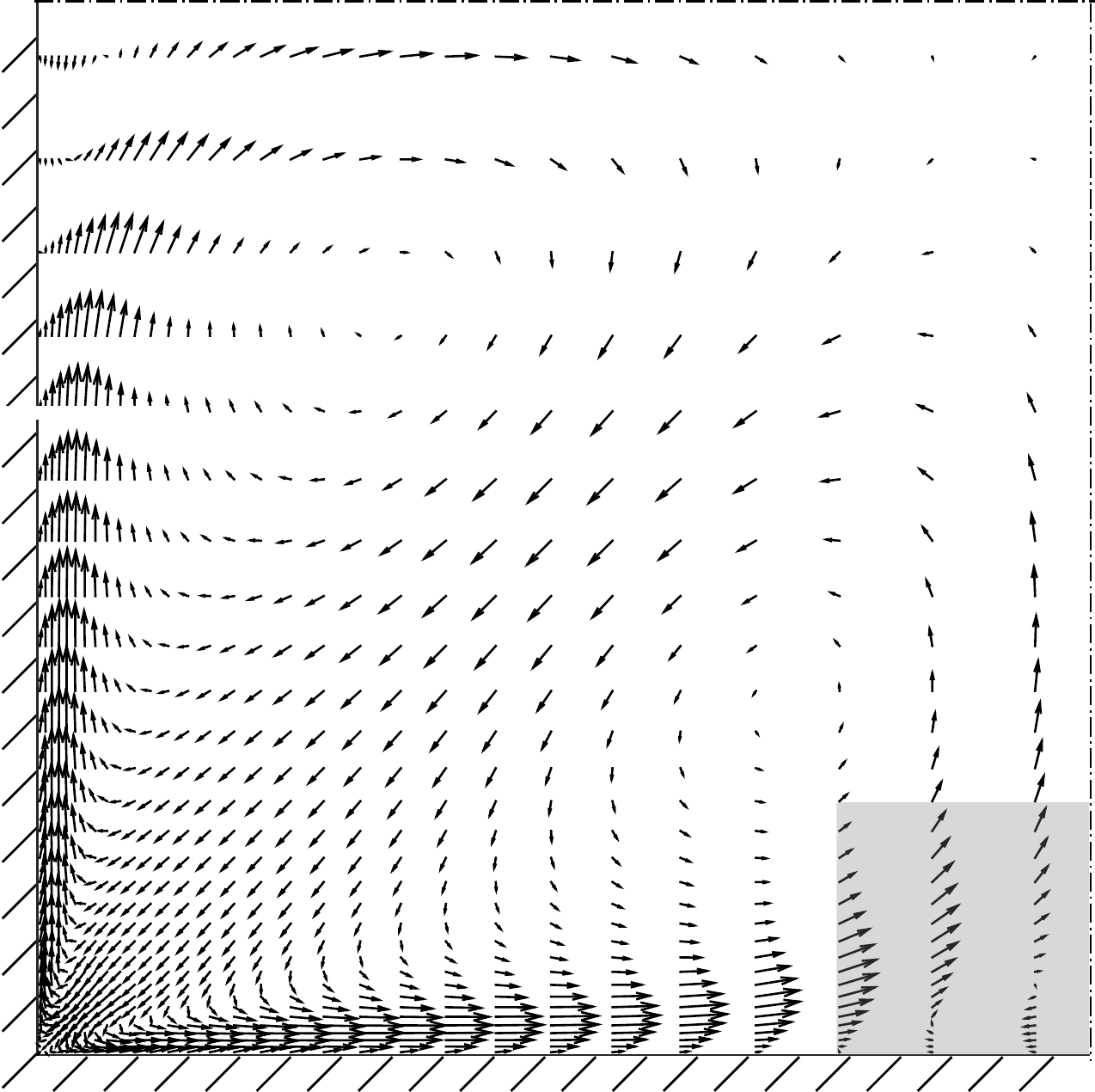}}\hspace{0.5em}
  \subfloat[Zoom-in view of separation]{\includegraphics[width=0.32\textwidth]{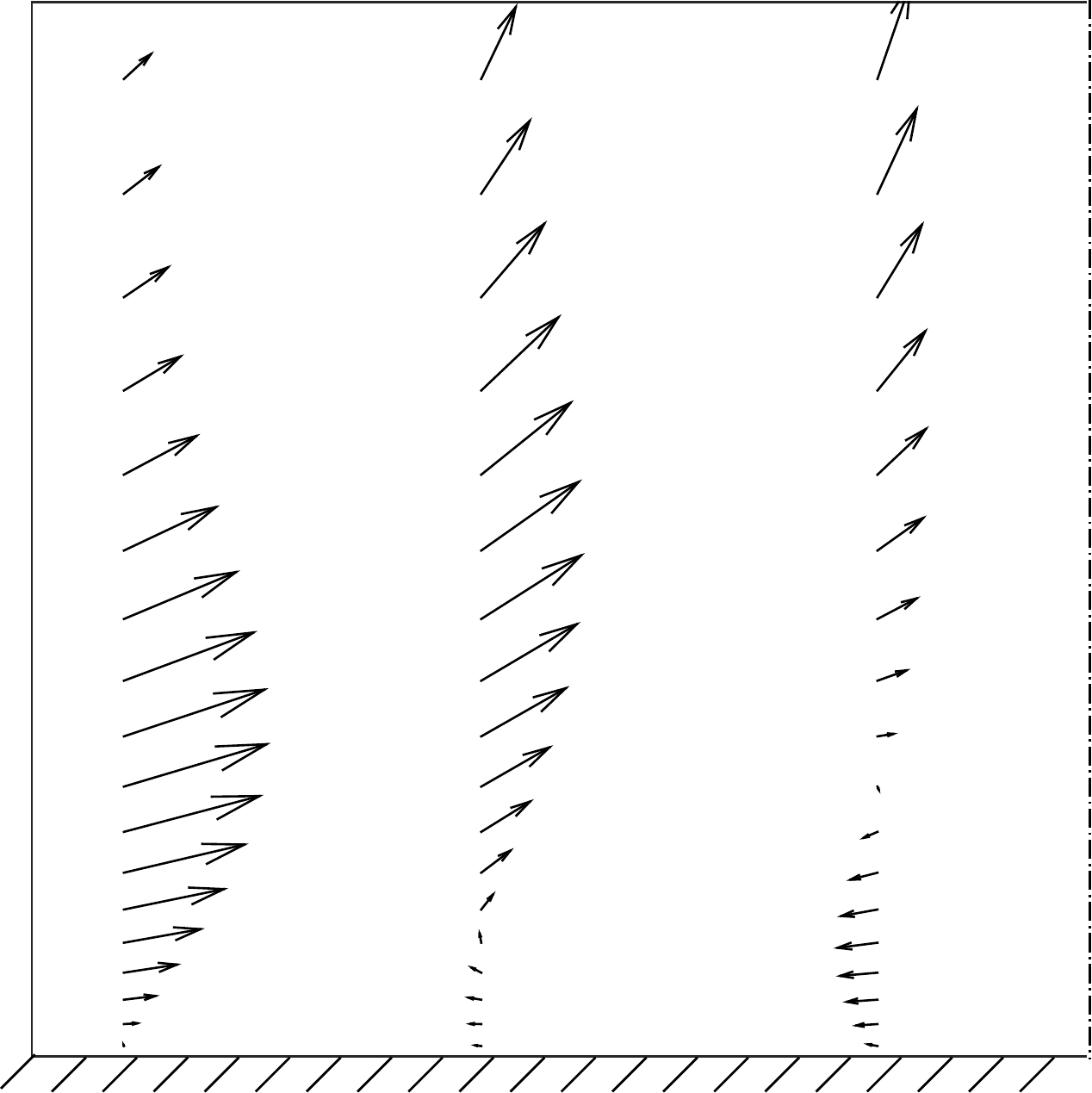}}\\
  \caption{Vector plot of the posterior mean of the predicted in-plane velocities for the flows at
    (a) $Re=1 \times 10^4$ (flow for calibration), (b) $Re=8.3 \times 10^4$ (flow to be predicted),
    (c) $Re=2.5 \times 10^5$ (flow to be predicted), and (d) the enlargement of shaded region in plot (c).  
    The directions and lengths of arrows
    indicate the directions and magnitude, respectively, of the secondary velocities. The shaded
    region highlight the presence of separation of secondary flow in the lower-left region (near the
    bottom wall and the vertical axis of symmetry).
\label{fig:vector_duct}
} 
\end{figure}

Figure~\ref{fig:vector_duct} shows vectors plots of the secondary flow velocity in the calibration
and prediction cases. The vortex structures in both flows, which are at Reynolds numbers $Re =8.3
\times 10^4$ and $Re = 2.5 \times 10^5$, are successfully captured as shown in
Figs.~\ref{fig:vector_duct}a and~\ref{fig:vector_duct}b, respectively. In addition,
  two general trends can be observed from Fig.~\ref{fig:vector_duct} as the Reynolds number
  increases.  First, the secondary flow penetrates further to the corner of the duct as the Reynolds
  number increases. That is, the corner region where secondary flow is absent becomes smaller with
  increasing Reynolds number.  Second, the thickness of the secondary flow boundary layers decreases
  as the Reynolds number increases, i.e., the velocity gradient $\partial U_y/ \partial z$ near the
  bottom wall increases as the flow Reynolds number increases.  Both trends are consistent with
  the findings from previous studies~\cite{brundrett64,madabhushi91}. Again, it is emphasized that
  these trends are predicted without utilizing observation data from the high Reynolds number cases
  ($Re=8.3\times 10^4$ or $Re=2.5 \times 10^5$). 

In addition to the favorable quantitative agreement with experimental data presented above, we also
found that the qualitative features of the flows at Reynolds numbers $Re=8.3 \times 10^4$ and $2.5
\times 10^5$ are captured well in the prediction, although full-field benchmark data are not
available for a detailed comparison.  Vector plots of the posterior mean of the calibrated and
predicted secondary velocities are presented in the three panels of Fig.~\ref{fig:vector_duct}.
Comparison of Figs.~\ref{fig:vector_duct}b and~\ref{fig:vector_duct}c shows that the predicted
overall flow patterns of the two flows at $Re=8.3 \times 10^4$ and $2.5 \times 10^5$ are very
similar except for the minor differences near the lower left corner and in the near wall
region. This is consistent with the previous findings reported in the
literature~\cite[e.g.,][]{brundrett64} that the general patterns of the secondary flows are not
sensitive to the increase of Reynolds number at high Reynolds numbers. A closer examination reveals
that the flow pattern does become more complex as the Reynolds number increases. For example, Figure~\ref{fig:vector_duct}c shows that in the flow at $Re=2.5\times10^5$ the secondary flow starts
to separate near the lower right part of the computational domain (the shaded region). This
prediction is physical and can be confirmed by the available experimental data presented in
Fig.~\ref{fig:Uy_Re10550}b, in which the secondary flow velocity $U_y$ becomes negative (leftward)
at the same location~\cite{gessner81}. Although such a phenomenon is not present in the calibration
flow at a lower Reynolds number $Re=1 \times 10^4$ (see Fig~\ref{fig:vector_duct}a), it is still
successfully captured by the proposed method for the flow at a higher Reynolds number $Re =
2.5\times10^5$.  Recall that in the present Bayesian framework, information in the prediction can
only come from the specified prior knowledge, observation data, or the physical dynamics model
(i.e., the RANS solver), but the former two sources do not have information on presence of the
separation. Therefore, one can only attribute the successful prediction of the separation to the
RANS solver. This finding clearly demonstrates the merits of fully utilizing the physical model in
the present method.

\subsection{Prediction of Flow in a Different Geometry}
\label{sec:res-geometry}
In Section~\ref{sec:pred-re} above, it has been demonstrated that the calibrated Reynolds stress
discrepancy can be utilized to predict flows on a geometrically similar domain but at higher Reynolds
numbers.  In this section we examine a more challenging case, where the Reynolds stress
discrepancies calibrated on the square duct flow are used to predict the flow in a rectangular duct
with aspect ratio of 1:2. The Reynolds number of the flow in the rectangular duct is $Re=8.8
\times 10^4$, which is moderately higher than that of the calibration flow ($1 \times
10^4$). However, we focus on the change of geometry here since the test cases above have
demonstrated that the change of Reynolds number does not pose intrinsic difficulties for the
extrapolation.

Three possible schemes have been proposed in Section~\ref{sec:extra-geo} to extrapolate the
calibrated Reynolds stress discrepancies on the square duct to the rectangular domain in the
prediction case.  The schemes have been illustrated in Fig.~\ref{fig:ext-geometry} and are
summarized below for completeness: (1) direct mapping of the square to the lower half of the
rectangle, (2) linear stretching of the square to the rectangle, and (3) direct mapping of the lower
triangle and linear stretching of the upper triangle.

The three seemingly arbitrary choices are in fact based on three clear assumptions of the
rectangular duct flow.  Comparing the domains of the calibration case in Fig.~\ref{fig:ext-geometry}
and the prediction case in Fig.~\ref{fig:ext-geometry} suggests that the major difference is that
the upper boundary CD is moved further away from the bottom wall AB.  The assumptions, i.e.,
assumed physical prior knowledge, on the change of the boundary location behind the three schemes
can be described as follows:
\begin{description}
\item[Scheme 1:] The change of the upper boundary location does not change the size and locations of the
  two vortices, and new flow patterns develop in the upper half of the rectangle.
\item[Scheme 2:] The flow in the rectangular duct maintains the same symmetric pattern as in the
  square duct case, with both vortices elongated approximately to the same aspect ratio of the
  rectangle.
\item[Scheme 3:] The change of the upper boundary location influences the upper vortex structure only, and the lower
  vortex remains unchanged.
\end{description}
While possibly none of the three assumptions is completely physical, the third one is the most
realistic, which we will justify as follows.  One of the most prominent features of the flows in both
the square and the rectangular ducts is the secondary flows induced by the anisotropy of Reynolds
stress tensor. This anisotropy is caused by the interactions between the boundary layers along the
two perpendicular walls (AD and AB).  It should be expected that moving the upper boundary further
away has larger influence on the boundary layer of the vertical wall (AD) than on that of the
horizontal wall (AB), since the flow in the upper half of the rectangular duct interacts only weakly
with the boundary layer of the bottom wall. Based on this reasoning, the assumption associated with
scheme 3 seems most reasonable, i.e., the change of boundary location influences only on the upper
vortex and no the lower vortex. Consequently, scheme 3 as shown in Fig.~\ref{fig:ext-geometry} is the
more appropriate choice.

\begin{figure}[htbp]
\centering
\subfloat[Experiment~\cite{hoagland60}]{\includegraphics[width=0.205\textwidth]{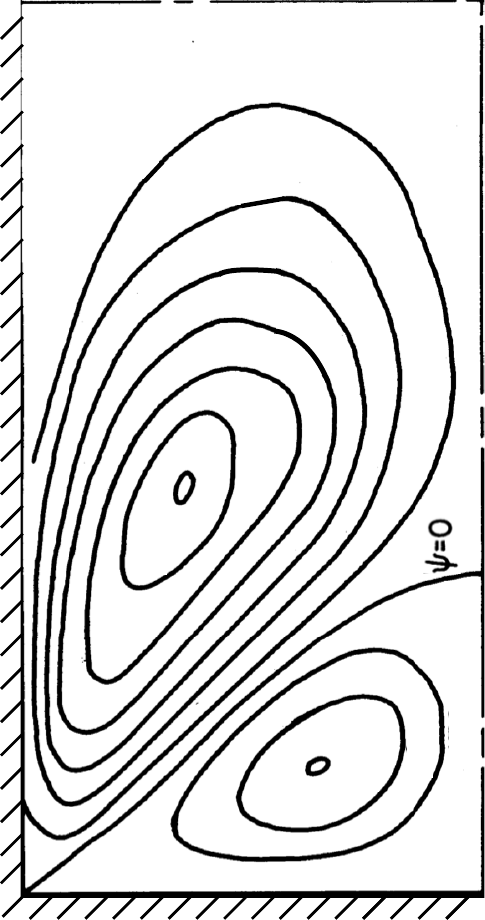}}\hspace{0.2em}
\subfloat[Scheme 1]{\includegraphics[width=0.2\textwidth]{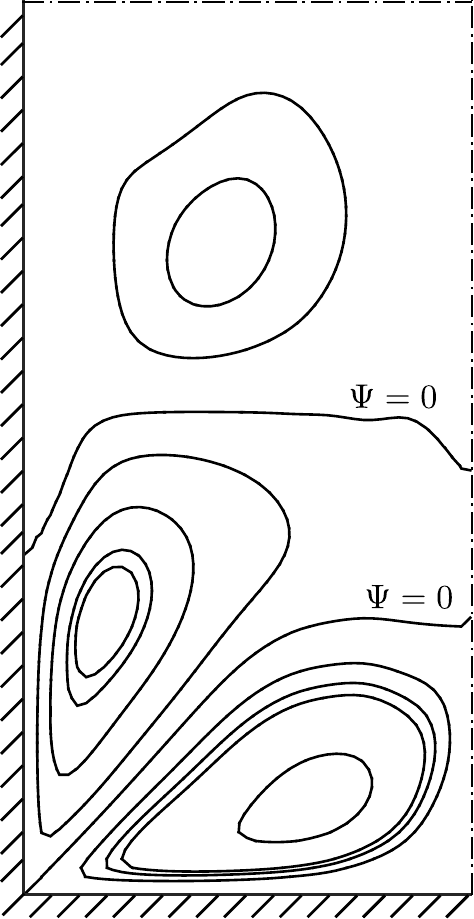}}\hspace{0.2em}
\subfloat[Scheme 2]{\includegraphics[width=0.2\textwidth]{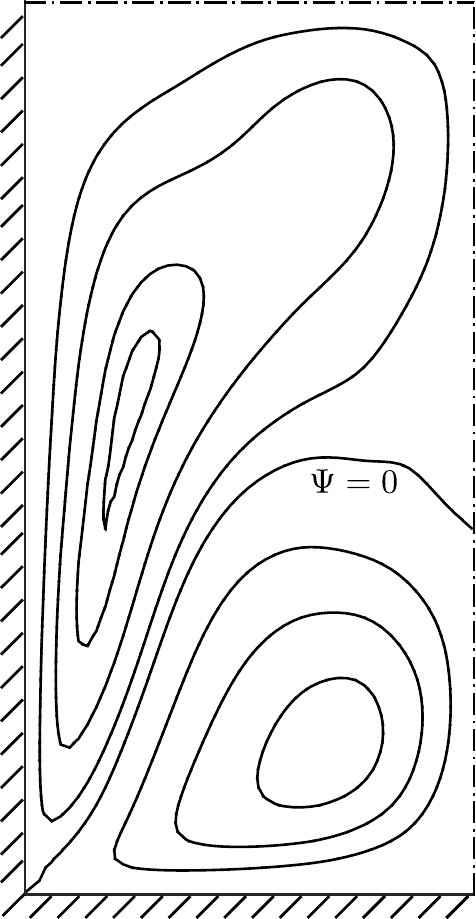}}\hspace{0.2em}
\subfloat[Scheme 3]{\includegraphics[width=0.2\textwidth]{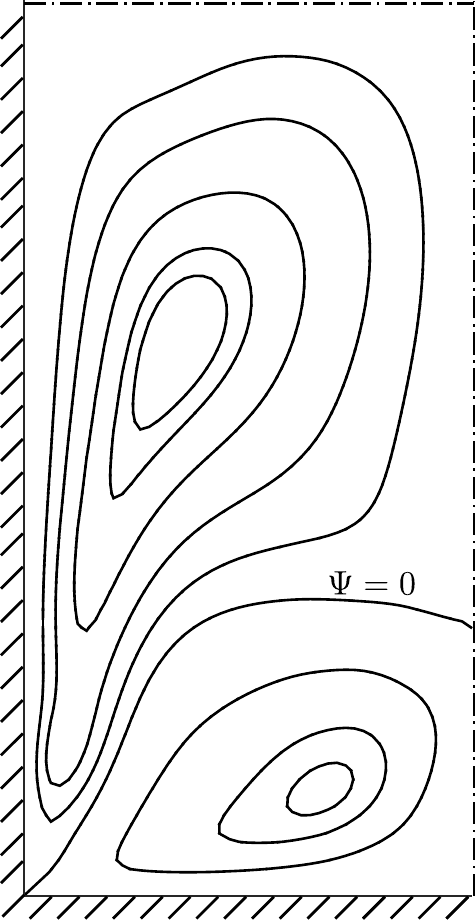}}
\caption{Comparison of experimentally measured and the predicted secondary flow patterns (posterior
  mean) based on three schemes of mapping Reynolds stress discrepancy from the calibration case to
  prediction case.  Panel (a) shows streamlines obtained from experiments~\cite{hoagland60}. Panels
  (b)--(d) show predicted streamlines based on schemes 1--3, respectively.  The baseline RANS
  prediction is omitted here as it predicts no secondary flows.
\label{fig:stream_comp}
} 
\end{figure}

The obtained flow patterns with the three mapping schemes above largely reflect their respective
physical assumptions as detailed above.  The in-plane streamlines from experiments~\cite{hoagland60}
are presented in Fig.~\ref{fig:stream_comp}a, which shows two vortices.  The two vortices in the
rectangular duct have different sizes, which is in contrast to the two symmetric vortices in the
square duct. The major vortex is located at the upper right, while the minor structure is located at
the lower left. The relative locations of the two vortices are similar to those in the square duct
flow.  The predicted streamlines based on the mapping schemes 1--3 are presented in
Fig.~\ref{fig:stream_comp}b--\ref{fig:stream_comp}d, respectively. Figure~\ref{fig:stream_comp}b
shows two vortices in the lower half of the duct similar to those in the square duct flow. It is
noteworthy that a third vortex develops on the upper half of the duct, which is driven by the vortex
below it. This predicted flow pattern is qualitatively different from that shown in
Fig.~\ref{fig:stream_comp}a, suggesting that the assumption associated with scheme~1 is unphysical.
Figure~\ref{fig:stream_comp}c shows that the prediction based on scheme~2 does capture the
qualitative features of the major and the minor vortices, although the predicted center locations
and the vortices sizes do not completely agree with the experimental results.  Finally, the
predicted streamlines based on scheme~3 are presented in Figure~\ref{fig:stream_comp}d, which has an
even better agreement with the experimental data compared to that obtained with scheme~2. In
particular, the center and the size of the minor vortex are very well predicted.  The general
pattern of the major vortex is also captured successfully, although the prediction quality is not as
good as for the minor vortex.  The remaining difference can be explained by the fact that the
mapping scheme is linear, which may not be true due to the different influences of the boundary layer
interactions.

Comparison of the three predictions above suggest that the assumed physical knowledge has a critical
impact on the obtained predictions.  An unphysical assumption as that behind scheme~1 leads to a
qualitatively incorrect prediction of overall flow patterns. An assumption based on sound physical
reasoning as in scheme~3 leads to the more favorable predictions.

\begin{figure}[htbp]
\centering
\subfloat[Experiment (Hoagland, 1960)]{\includegraphics[width=0.5\textwidth]{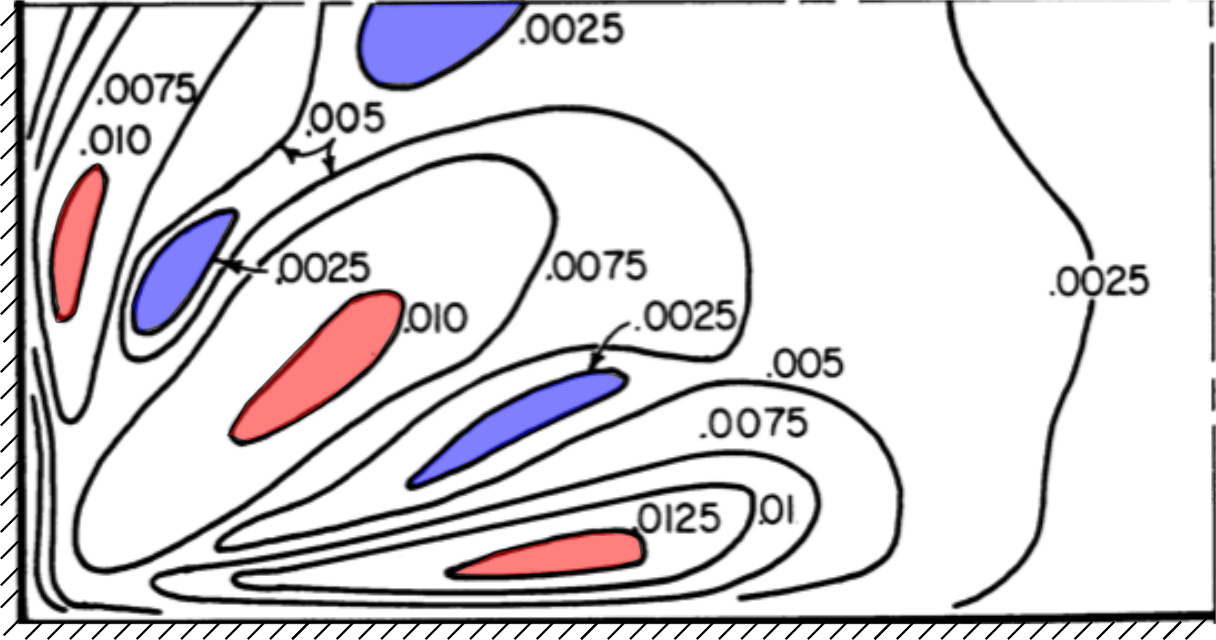}}\\
\subfloat[Prediction]{\includegraphics[width=0.5\textwidth]{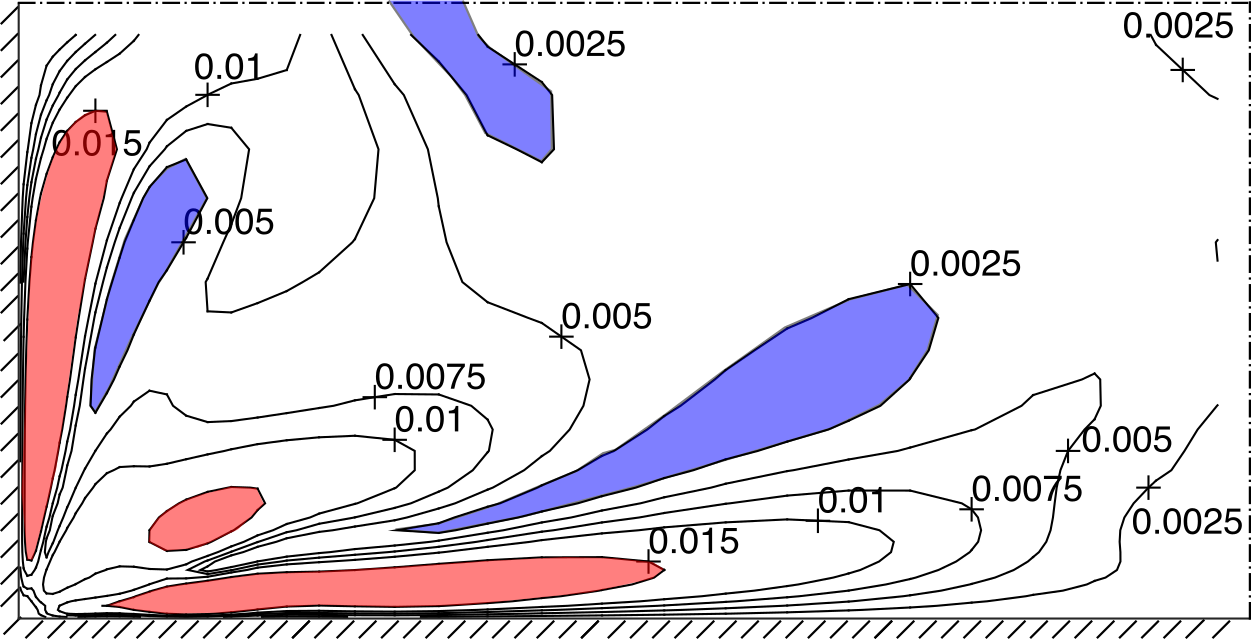}}
\caption{Comparison of experimentally measured and the predicted contours of secondary velocity
  magnitude (posterior mean). Panel (a) shows contours obtained from
  experiment~\cite{hoagland60}. Panel (b) shows predicted contours based on scheme 3.  Black crosses
  (+) and the numbers nearby indicate the values of the contours. Peaks and troughs are denoted as
   red (light shade) and blue (dark shade) to facilitate interpretation of flow field structure. 
   The baseline RANS prediction is omitted here as it predicts no secondary flows.}
\label{fig:vsec_comp}
\end{figure}

 Figure~\ref{fig:vsec_comp} shows the comparison of secondary flow velocity
  magnitude contours between experimental data~\cite{hoagland60} and the prediction based on
  scheme~3. It can be seen that the predicted contour patterns are similar to those of the
  experimental data, and the predicted magnitude is also comparable to the experimental
  data. Admittedly, the predicted secondary flow field is not exactly the same as the experimental
  data, e.g., the velocity magnitude is greater than the experimental data at the near wall region
  and smaller than the experimental data away from the wall. A possible reason is that the prior
  physical knowledge incorporated in scheme~3 is still far from enough to accurately account for
  the additional physics introduced by the geometry change. However, it can be seen from
  Fig.~\ref{fig:stream_comp} that the prediction qualitatively captures the flow pattern, which is
  totally absent in baseline RANS results, with much lower computational cost compared to high
  fidelity simulations. Therefore, we argue that the proposed method provides a practical approach
  to quickly and approximately search the design space with low computational costs in the
  preliminary design stage of engineering systems. It will help identify the promising design
  candidates for the further investigations with high fidelity simulations and/or model
  experiments.

\section{Conclusion}
\label{sec:conclusion}

Recently, Xiao et al.~\cite{xiao-mfu} proposed a framework for quantifying and reducing uncertainty
in RANS simulations based on sparse observations of velocities. In the present work we extend the
original framework to flows with no observation data, leading to a Bayesian calibration--prediction
method.  As in the original framework, the model-form uncertainty in RANS simulations are localized
to the Reynolds stresses, which are modeled as a random field. The uncertainty distribution of the
discrepancy is first calibrated with available data on a related flow (e.g., the flow in a
geometrically similar domain or in a slightly different geometry). Subsequently, the obtained
distribution is extrapolated to the prediction case and is sampled to correct the RANS-modeled
Reynolds stresses. The merits of the proposed method are demonstrated with two canonical flows of
engineering relevance, the flow over periodic hills and the flow in a square duct.  In the case of
periodic hill flows, the uncertainty distribution of the discrepancy is calibrated in the flow at a
lower Reynolds number and extrapolated to a higher Reynolds number. In the duct flow cases, the
discrepancy distribution is calibrated in a square duct flow and used to predict flows at higher
Reynolds number flows and that in a rectangular duct.  Numerical simulation results demonstrate that
the predictions of posterior mean velocities have significantly improved agreement with the
benchmark data compared to the baseline RANS predictions.  It is noteworthy that even the flow
features that are not present in the calibration flow have been successfully captured in the
prediction case by the proposed procedure, demonstrating the merits of framework in fully utilizing
the RANS model.  Based on the simulation results, we conclude that the proposed
calibration--prediction method is a promising candidate for reducing model-form uncertainties in
RANS simulations.

The proposed method is an algorithmically straightforward yet practically profound extension of the
original framework.  Since direct observation data are not required on the flow to be predicted, the
range of applications is greatly expanded compared to the original framework. This is particularly
relevant for turbulent flow simulations in support of engineering design, where the prototype system
has not been build yet in the design stage.

An important assumption in the proposed method is that the flow used for calibration and the flow to
be predicted are closely related. Specifically, they share the same overall characteristics despite
the fact that the two flows are not dynamically similar (i.e., they have different Reynolds numbers)
or not even geometrically similar (e.g., flow in a square duct versus flow in a rectangular duct).
This assumption is valid in many scenarios in engineering design. On the other hand, the assumption
also implies that the quality of the prediction inevitably depends on the judgment of the
analyst. We argue that this is an intrinsic feature of Bayesian methods, and it is often an
advantage for a framework to be able to incorporate insights and prior knowledge from the users, who
are often experts of the model and the problem of concern.



\bibliographystyle{elsarticle-num}
\bibliography{mfu3,../1-general-methodology/career,../1-general-methodology/mfu,../1-general-methodology/DOE}

\appendix 

\section{Algorithm for Resampling the Posterior Distribution of Reynolds Stress Discrepancies}
 \label{sec:resample}

 Given an ensemble $\{\bs{\omega}_i\}_{i=1}^N$ of size $N$ representing the posterior distribution
 of Reynolds stress discrepancy, the objective is to obtain another ensemble of size $N'$ drawn from
 the same distribution.  Each sample $\bs{\omega}_i$ is a vector of size $3m \times 1$, consisting
 of the coefficients for the retained $m$ modes of the three discrepancy fields $\delta^k$,
 $\delta^\xi$ and $\delta^{\eta}$.  The following algorithms are used for the resampling:
\begin{enumerate}
\item Estimate the sample mean and sample co-variance of the ensemble $\{\bs{\omega}_i\}_{i=1}^N$ as
  follows:
  \begin{align}
    \bar{\bs{\omega}} & = \frac{1}{N}\sum_{i = 1}^{N} \bs{\omega_i}   \label{eq:sample-mean} \\
    C & =  \frac{1}{N-1} \sum_{i=1}^N (\bs{\omega}_i - \bar{\bs{\omega}} )^T  (\bs{\omega}_i -
    \bar{\bs{\omega}} )  \label{eq:sample-cov}
  \end{align}
\item Perform eigendecomposition of covariance matrix $C$ as below
  \begin{equation}
    \label{eq:eigen}
    C \phi_j = \lambda'_j  \phi_j 
  \end{equation}
  to obtain the orthogonal eigenvectors $\{ \phi_j \}_{j = 1}^{3m}$. 
\item Project the samples $\{\bs{\omega}_i\}_{i=1}^N$ on the basis functions $\{ \phi_j \}_{j =
    1}^{3m}$ to obtain $\{ \bs{\alpha}_i \}_{i=1}^N$, where $\bs{\alpha}_i$ is the coordinate of the
  sample $\bs{\omega}_i$ on the new basis set $\{ \phi_j \}_{j = 1}^{3m}$.  The transformed ensemble
  $\{ \bs{\alpha}_i \}_{i=1}^N$ can be considered realizations of a random vector $\mathbf{a}$ of
  size $3m \times 1$ with uncorrelated components.
\item The probability distribution functions and the corresponding cumulative density functions
  (CDF) for each component of the random vector $\mathbf{a}$ is estimated from the samples, i.e.,
  $\{ \bs{\alpha}_i \}_{i=1}^N$, by using kernel density estimation techniques.
\item Generate $N'$ samples for each of the component from the estimated CFD by using standard
  sampling techniques~\cite{Glasserman:2004ua}.
\item Reconstruct sample $\{ \bs{\omega}'_i \}_{i=1}^{N'}$ of size $N'$ from the resampled 
coefficients $\{ \bs{\alpha}'_i \}_{i=1}^N$ and the basis $\{ \phi_j \}_{j = 1}^{3m}$.
\end{enumerate}

\end{document}

%% file: headers/header-common.tex
\usepackage{graphicx}
\usepackage{amsmath}

\usepackage{natbib}
\usepackage{amssymb}

\usepackage{amsthm}

\usepackage{lineno}
\usepackage{subfig}
\usepackage{enumerate}
\usepackage{fullpage}
\usepackage{algorithm}
\usepackage{algpseudocode}
\usepackage{xcolor}
\usepackage[colorinlistoftodos]{todonotes}

\usepackage{hyperref}

\biboptions{sort,compress} 

\usepackage{xcolor}

\newcommand*\patchAmsMathEnvironmentForLineno[1]{%
  \expandafter\let\csname old#1\expandafter\endcsname\csname #1\endcsname
  \expandafter\let\csname oldend#1\expandafter\endcsname\csname end#1\endcsname
  \renewenvironment{#1}%
     {\linenomath\csname old#1\endcsname}%
     {\csname oldend#1\endcsname\endlinenomath}}%
\newcommand*\patchBothAmsMathEnvironmentsForLineno[1]{%
  \patchAmsMathEnvironmentForLineno{#1}%
  \patchAmsMathEnvironmentForLineno{#1*}}%
\AtBeginDocument{%
\patchBothAmsMathEnvironmentsForLineno{equation}%
\patchBothAmsMathEnvironmentsForLineno{align}%
\patchBothAmsMathEnvironmentsForLineno{flalign}%
\patchBothAmsMathEnvironmentsForLineno{alignat}%
\patchBothAmsMathEnvironmentsForLineno{gather}%
\patchBothAmsMathEnvironmentsForLineno{multline}%
}


%% file: headers/header-xiao.tex
\usepackage{color,soul}
\usepackage{enumitem}
\usepackage{mathtools}
\usepackage{booktabs}

\definecolor{lightblue}{rgb}{.90,.95,1}
\definecolor{darkgreen}{rgb}{0,.5,0.5}

\definecolor{lightgreen}{rgb}{.90,1,0.90}

\newcommand{\bstau}{\boldsymbol{\tau}}
\newcommand{\bstaurans}{\tilde{\boldsymbol{\tau}}^{rans}}
\newcommand{\bs}[1]{\boldsymbol{#1}}
